\documentclass[a4paper]{article}
\usepackage[margin=1.0in]{geometry}
\usepackage[dvipsnames]{xcolor}
\usepackage{amsmath, amsthm, amsfonts, amssymb, xcolor, multirow, slashed, soul}
\usepackage{jheppub}
\usepackage{physics}
\usepackage[table]{}
\usepackage[capitalize]{cleveref}
  \crefname{section}{Sec.}{Secs.}

\hypersetup{
    linktocpage=true,
    colorlinks=true,
    linkcolor=blue,
    citecolor=red,
    pdftitle={Title},
    bookmarks=true,
    pdfpagemode=FullScreen,
}

\usepackage[shortlabels,inline]{enumitem}	
\usepackage[capitalize]{cleveref}	
  \crefname{section}{Sec.}{Secs.}
  \crefname{appendix}{App.}{Apps.}
  \crefname{enumi}{case}{cases}

\newcommand{\be}{\begin{equation}}
\newcommand{\ee}{\end{equation}}
\newcommand{\bea}{\begin{equation}\begin{aligned}}
\newcommand{\eea}{\end{aligned}\end{equation}}

\providecommand{\oper}{\mathcal{O}}
\providecommand{\ord}{\mathcal{O}}
\providecommand{\ie}{i.e.}
\providecommand{\eg}{e.g.}

\DeclareMathOperator{\Disc}{Disc}
\newcommand{\amp}{\mathcal{A}}

\renewcommand{\vec}[1]{\vb{#1}}  
\newcommand{\vk}{\vec{k}}

\newcommand{\cs}{c_{\rm s}}

\providecommand{\ZZ}{\mathbb{Z}}

\newcommand{\Mpl}{M_\mathrm{pl}}
\newcommand{\dpi}{\dot{\pi}}
\newcommand{\dpic}{\dot{\pi}_\mathrm{c}}
\newcommand{\dsig}{\dot{\sigma}}

\newcommand{\delg}{\delta g^{00}}

\newcommand{\pid}{\dot{\pi}}
\newcommand{\pidc}{\dot{\pi}_\mathrm{c}}

\newcommand{\delpic}{\partial\pi_\mathrm{c}}
\newcommand{\delipi}{\partial_i\pi}
\newcommand{\delpisq}{(\partial\pi)^2}
\newcommand{\delpisqc}{(\partial\pi_\mathrm{c})^2}

\newcommand{\phid}{\dot{\phi}_0}

\newcommand{\sigd}{\dot{\sigma}}
\newcommand{\delsigma}{\partial\sigma}
\newcommand{\delisigma}{\partial_i\sigma}
\newcommand{\delsigsq}{(\partial\sigma)^2}

\newcommand{\bk}{\mathbf{k}}

\newcommand{\Refc}[1]{Ref.~\cite{#1}}

\begin{document}
\preprint{CERN-TH-2022-160}

\title{\LARGE{\textbf{Multifield Positivity Bounds for Inflation}}
}

\author[a]{Marat Freytsis,}
\author[b, c]{Soubhik Kumar,}
\author[d,e]{Grant N. Remmen,}
\author[f]{Nicholas L. Rodd}

\affiliation[a]{NHETC, Department of Physics and Astronomy, Rutgers University, Piscataway, NJ 08854, USA}
\affiliation[b]{Berkeley Center for Theoretical Physics, Department of Physics, University of California, Berkeley, CA 94720, USA}
\affiliation[c]{Theoretical Physics Group, Lawrence Berkeley National Laboratory, Berkeley, CA 94720, USA}
\affiliation[d]{Kavli Institute for Theoretical Physics,
University of California, Santa Barbara, CA 93106, USA}
\affiliation[e]{Department of Physics,
University of California, Santa Barbara, CA 93106, USA}
\affiliation[f]{CERN, Theoretical Physics Department, Geneva 1211, Switzerland}
\emailAdd{marat.freytsis@rutgers.edu}
\emailAdd{soubhik@berkeley.edu}
\emailAdd{remmen@kitp.ucsb.edu}
\emailAdd{nrodd@cern.ch}

\abstract{Positivity bounds represent nontrivial limitations on effective field theories (EFTs) if those EFTs are to be completed into a Lorentz-invariant, causal, local, and unitary framework.
While such positivity bounds have been applied in a wide array of physical contexts to obtain useful constraints, their application to inflationary EFTs is subtle since Lorentz invariance is spontaneously broken during cosmic inflation.
One path forward is to employ a \emph{Breit parameterization} to ensure a crossing-symmetric and analytic S-matrix in theories with broken boosts.
We extend this approach to a theory with multiple fields, and uncover a fundamental  obstruction that arises unless all fields obey a dispersion relation that is approximately lightlike.
We then apply the formalism to various classes of inflationary EFTs, with and without isocurvature perturbations, and employ this parameterization to derive new positivity bounds on such EFTs.
For multifield inflation, we also consider bounds originating from the generalized optical theorem and demonstrate how these can give rise to stronger constraints on EFTs compared to constraints from traditional elastic positivity bounds alone.
We compute various shapes of non-Gaussianity (NG), involving both adiabatic and isocurvature perturbations, and show how the observational parameter space controlling the strength of NG can be constrained by our bounds.
}

	
\setcounter{footnote}{0}
\maketitle



\section{Introduction}

Effective field theories (EFTs) represent a modern, powerful method for efficiently specifying the degrees of freedom and interactions relevant in some finite kinematic range in a way that is both systematic and systematically improvable\cite{Fermi:1934sk,Heisenberg:1936nmg,Weinberg:1978kz,Caswell:1985ui,Manohar:2000dt,Polchinski:1992ed,Bauer:2000yr,Goldberger:2004jt}.
Upon identifying the relevant external degrees of freedom, one first writes down all the operators that are consistent with the symmetries.
These operators capture the effects of degrees of freedom that \emph{cannot} be accessed as external, dynamical states, such as heavy particles.
To obtain physical predictions, one truncates to a finite subset of operators by implementing a power counting rule for these operators in one or more small parameters---such as a derivative expansion in units of the ultraviolet (UV) scale---with the power-counting expansion providing the required number of operators to reach a given precision.

This Wilsonian EFT approach is very general and agnostic of the UV description of the system.
One might then naively think that any choice of parameters for the EFT operators---the so-called Wilson coefficients---corresponds to some family of UV completions.
However, this is not the case, and certain choices of Wilson coefficients can be ruled out for any EFT descended from a healthy UV that is causal, local, unitary, and Lorentz invariant~\cite{Adams:2006sv}.
In a Lorentz-invariant EFT, such \emph{positivity bounds} require the forward elastic scattering amplitude $\amp(s)$ to obey $\amp''(s \to 0) > 0$,\footnote{Throughout, we will use {\it forward} to mean equality of incoming and outgoing momenta for particles 1 and 3, and similarly for particles 2 and 4, irrespective of any additional quantum numbers.
Requiring that the total incoming two-particle state be identical to the outgoing state---both in momentum and in terms of internal quantum numbers like flavor---will be referred to as both forward and {\it elastic}.} with the prime indicating a derivative with respect to $s$.

This requirement can directly constrain signs of various quartic dimension-eight operators in an EFT.
A classic example is the operator $\mathcal{L} \supset c\, (\partial\phi)^4 /\Lambda^4$ in a scalar EFT~\cite{Adams:2006sv}.
While from a purely bottom-up perspective $c$ can take either sign, demanding that the operator arises from a healthy UV theory forces $c>0$.
This idea of imposing positivity bounds on EFT coefficients has been implemented in a variety of physical scenarios, with early works~\cite{Pham:1985cr,Ananthanarayan:1994hf,Pennington:1994kc} in the context of chiral perturbation theory. 
For a recent review of these ideas, see \Refc{deRham:2022hpx}.

Inflationary EFTs represent a particularly interesting set of candidates from the point of view of positivity bounds for a number of reasons.
First, during inflation time translation invariance is spontaneously broken by the inflaton background.
Therefore, in the infrared (IR) at energies of order the inflationary Hubble scale $H$, one cannot directly use a Lorentz-invariant description of the system.
Consequently, two-to-two scattering amplitudes depend on additional kinematic variables beyond the standard Mandelstam $s,t,u$ variables.
Unless these amplitudes are parameterized properly, the additional kinematic dependencies give rise to spurious nonanalyticities of the S-matrix.
Similar considerations apply to other EFTs that break boosts, such as that describing a superfluid.
Therefore, the study of positivity bounds on inflationary EFTs can serve as a paradigmatic example for a wider class of EFTs.

Second, the observables giving access to inflationary dynamics are the non-Gaussianity (NG) of primordial density perturbations.
Such NG contributions are determined by the \emph{expectation values} of (iso)curvature perturbations at the end of inflation.
Owing to this fact, within the various EFTs of inflation discussed above NG can receive contributions at linear order in their coefficients.
At the same time, positivity bounds computed from scattering amplitudes can also involve linear relationships among EFT coefficients.
In this sense, a direct connection exists between the parameter space probed by NG measurements and positivity bounds.
By imposing such bounds, it can also be checked whether certain parameter regions or NG configurations that are difficult to probe observationally are in fact forbidden.
This linearity stands in contrast with observables that are sensitive only to the squares of matrix amplitudes, such as cross sections.
Such observables depend quadratically on the underlying Wilson coefficients, and accessing information related to positivity usually requires the study of interference effects.

In this work, we consider four classes of inflationary EFTs.
\begin{enumerate}[(a)]
	\item\label{item:gEFT}
	      \hspace{-0.5em} The Goldstone EFT of inflation~\cite{Cheung:2007st};
	\item\label{item:multadaEFT}
	      \hspace{-0.5em} An EFT of multifield inflation~\cite{Senatore:2010wk} with a single effective degree of freedom leading to adiabatic perturbations;
	\item\label{item:multisoEFT}
	      \hspace{-0.5em} An EFT of multifield inflation~\cite{Senatore:2010wk} leading to isocurvature perturbations; and
	\item\label{item:srEFT}
	      \hspace{-0.5em} A Lorentz-invariant EFT of multifield inflation in the slow-roll regime, also giving rise to isocurvature perturbations.
\end{enumerate}
We clarify that for all the ``multifield'' scenarios mentioned above, we will be considering situations where the homogeneous inflationary expansion is still driven only by the inflaton field, while perturbations come from fluctuations of both the inflaton field and a second \emph{spectator} field.
This is thus different from the multifield scenarios where multiple fields drive the inflationary expansion itself.
However, in the case of massless multifield inflation, we can without loss of generality rotate the field definitions such that only a single field sources the vev.

For \cref{item:gEFT,item:multadaEFT,item:multisoEFT}, Lorentz invariance is broken spontaneously at the energy scales of interest, and subtleties involving additional kinematic dependence of the S-matrix can arise, as mentioned above.
To address these, we employ the \emph{Breit parameterization} of \Refc{Grall:2021xxm}.
This parameterization prescribes a way of writing down scattering amplitudes such that, in principle, they are crossing symmetric and analytic everywhere in the complex plane except the real $s$ axis, even when boosts have been broken.
Ref.~\cite{Grall:2021xxm} studied this formalism in theories that involve a single field.
Here we will extend the approach to two-field scenarios, where a novel complication is that the two fields can propagate with different speeds of sound $\cs$.
We will show explicitly that, when applying the Breit parameterization to such theories, additional nonanalyticities can arise in the form of extra poles both on and off the real $s$ axis.
The presence of complex poles undermines the conventional dispersive arguments on which positivity is based.
Even when the additional pole is on the real $s$ axis, we find that it comes with an opposite sign for the residue such that the standard dispersive arguments cannot be applied directly to claim $\amp''(s \to 0)>0$.
Therefore, throughout we will restrict our analysis to cases where all the propagation speeds are parametrically close to the speed of light.
We emphasize that even when $\cs \simeq 1$, we have not merely reduced the problem back to the conventional Lorentz-invariant scenario.
The existence of a preferred timelike direction allows for operators such as $\dot{\pi}^3$, where $\pi$ is a fluctuation and the overdot denotes a time derivative with respect to the preferred frame defined by the inflaton, to appear in the Lagrangian of the EFT whose presence would be forbidden in a Lorentz-invariant theory.

Along with elastic positivity bounds, we also consider bounds from the generalized optical theorem.
In particular, we consider two-field scenarios and demonstrate that the bounds from the generalized optical theorem can be more constraining compared to standard positivity bounds alone.
For inflationary scenarios, more than one degree of freedom can be present if there are surviving isocurvature perturbations.
Therefore, as an example, we consider the EFT for \cref{item:srEFT} described above and derive the associated generalized optical theorem bounds.
We also consider a special case with a dihedral symmetry group relating various EFT coefficients to draw a straightforward comparison between elastic positivity bounds and those from the generalized optical theorem.

The bounds we derive in this work further constrains the parameter space currently allowed by observations.
We demonstrate this explicitly using the EFT for \cref{item:multadaEFT} mentioned above.
As we will describe, in the presence of a $\ZZ_2$ symmetry, such an EFT can dominantly give rise to a nonzero four-point function (trispectrum) of curvature perturbations.
We can then impose the derived positivity bounds on the parameter space allowed by Planck searches for the primordial trispectrum~\cite{Planck:2019kim}.
Converting our bounds on the EFT in \cref{item:srEFT}---where we apply the generalized optical theorem---to the observational parameter space is more challenging.
The primary difficulty arises because the current searches for isocurvature NG assume the so-called local shape template~\cite{Planck:2019kim} that peaks at \emph{squeezed} momentum configurations.
On the other hand, the derivative interactions implied by EFT for \cref{item:srEFT}, for example, give rise to isocurvature NG peaking at \emph{equilateral} momentum configurations.
If an optimal search for these latter types of isocurvature NG with an appropriate template is performed in the future, then we can readily impose the derived bounds from analyticity and the generalized optical theorem on the resulting parameter space.
To that end, we have derived the shapes of various three-point functions in terms of the EFT coefficients.

As a final introductory remark, a few words are merited regarding a subtlety of what we mean by analytic continuation of an amplitude from low to high energies, and the assumptions associated with this implicit in our results.
When considering a forward amplitude ${\cal A}(s)$ with $s$ ranging from UV to IR scales, one can be concerned that the degrees of freedom in the EFT used to define the asymptotic states that are scattered could vary with scale, for instance as we move through scales associated with symmetry breaking or the transition to strong coupling.
Indeed, there is even a degree of ambiguity for a well defined theory: is the high energy equivalent of a quark in QCD also a quark, or instead a quark with a soft dressing of gluons?
Ultimately, amplitude type arguments which connect the UV and IR implicitly assume there is a well defined map in the Hilbert space from the one-particle degrees of freedom used to define ${\cal A}(s)$ in the IR to states in the UV from which we can still construct a $2 \to 2$ amplitude.
In theories with Lorentz invariance, the argument is on firmer footing as we can always boost the IR states to high energy, which suggests there exists a well defined path through the Hilbert space.
Without boosts, however, the situation is less clear and in principle the IR amplitude could map to an $m \to n$ amplitude in the UV, for which the usual assumptions invoked in dispersive arguments may fail.
In any case, for reasons we outline in Sec.~\ref{sec:twofield}, even in Lorentz-breaking EFTs, we will be taking external particles for which the speed of sound is parametrically close to unity, for which the boost of the external states themselves from the IR to the UV is less ambiguous.

For the above reasons, the results in the present work will always be contingent on either, 1) the existence of an IR to UV mapping between the external states sufficient to define a $2 \to 2$ amplitude in either limit; or 2) the appropriate amplitude we do map to in the UV is sufficiently well behaved for the conventional dispersive arguments to hold.
To expand on the second possibility, in order for a Froissart type relation to hold, all that we need is for the amplitude in the UV to scale as $s^{2-\epsilon}$ for $\epsilon>0$, or a small softening of the amplitude in the UV.
Given the requirement of the above assumption, there is also motivation to explore what positivity bounds can be derived from completely different starting points, such as the assumption that the UV flows to a conformal field theory, an approach that has been developed in \Refc{Creminelli:2022onn}.

This work is organized as follows.
In \cref{sec:twofield} we discuss how to implement positivity bounds on theories with spontaneously broken Lorentz invariance that contain two light degrees of freedom.
Along the way, we discuss aspects of the Breit parameterization and the novel challenges multiple fields introduce.
We apply the generalized optical theorem to write a general set of relations among various scattering amplitudes in such two-field scenarios in \cref{sec:gen_opt_thm}.
In \cref{sec:infeft}, we describe in detail the four classes of inflationary EFTs that we consider in this work.
We use the generalized positivity bounds to constrain these EFTs in \cref{sec:bounds} and map them to the observational parameter space in \cref{sec:obs}.
We discuss future directions and conclude in \cref{sec:concl}.
\Cref{app:bispec_comp} contains some of the details of the computation of three-point functions involving adiabatic and isocurvature perturbations.

\paragraph{Notations and conventions.}
Our metric convention will be mostly plus, \ie, $(-,+,+,+)$.
We will use natural units, $c = \hbar = 1$, and the reduced Planck mass, $\Mpl^2 \equiv 1/(8\pi G_N)$, where $G_N$ is the Newton constant.
We denote the four-momentum components of particle $i$ as $k_i^\mu \equiv (\omega_i,\vk_i)$.
When considering the breaking of Lorentz symmetry, we parameterize the breaking with a single timelike four-vector $n^\mu$ as in \Refc{Grall:2021xxm}.
The frame in which $n^\mu = (1, \vec{0})$ is the one in which quantities will be invariant under rotations and spatial translations and the one in which we will implicitly always discuss specific energy and momentum configurations.\footnote{We could instead insist on working with the quantities $k_i \cdot n$ and $k_i^\mu - (k_i \cdot n) n^\mu$, corresponding to $\omega_i$ and $(0, \vk_i)$, respectively, in the $n^\mu = (1, \vec{0})$ frame. This would allow us to phrase all results in a frame-independent manner, at the cost of making our expressions less transparent.}
The most general dispersion relation allowed under these assumptions is
\begin{equation}
  \omega_i^2 = \cs(\omega_i)^2 \abs{\vk_i}^2 + m(\omega_i)^2,
\end{equation}
with $\cs(\omega_i) \le 1$.
For the explicit UV completions we will consider, it will be sufficient to take the simpler case of $\cs$ and $m$ being constant and independent of $\omega_i$ (and hence $\vk_i$), in which case $\cs$ can be taken to be the speed of sound for the nonrelativistic field under consideration.
For most discussions, we will also work in the massless limit, $\omega \gg m$, so the dispersion relation will further simplify to $\omega_i = \cs \abs{\vk_i}$.
This simplification is well motivated since we will be studying the scattering of the inflaton and other light scalars at subhorizon scales with $s\gg H^2$.

Throughout this work, we will be interested in $2 \to 2$ scattering amplitudes with the all-incoming momentum convention.
We use the noncyclic labeling $(1243)$ such that $k_{3,4}^\mu \to -k_{1,2}^\mu$ corresponds to taking the forward limit.
As shorthand, we define $\omega_{ij} \equiv \omega_i + \omega_j$ and $s_{ij} \equiv \omega_{ij}^2 - \cs^2 |\vk_i+\vk_j|^2$.
In the relativistic limit, on-shell amplitudes are functions of Mandelstam variables $s \equiv s_{12} = -(k_1 + k_2)^2$ and $t \equiv s_{13} = -(k_1 + k_3)^2$ only.
On the other hand, in the boost-breaking case with preserved rotational and translational symmetry described above, amplitudes depend upon five variables, which we take to be $\omega_s \equiv \omega_{12}$, $\omega_t \equiv \omega_{13}$, $\omega_u\equiv\omega_{14}$, $s$, and $t$.
The Mandelstam variable $u$ is defined as $u \equiv s_{14} = -(k_1+k_4)^2$.
In the massless limit of both the Lorentz-invariant and -violating cases, we have $s+t+u=0$.

\section{Two-Field Lorentz-Violating Theories}
\label{sec:twofield}

In a theory violating Lorentz invariance, the scattering amplitudes are frame-dependent and are written in terms of additional kinematic variables beyond the three Mandelstam  $s, t, u$.
The additional variables can give rise to nonanalyticities of amplitudes away from the real $s$ axis, invalidating the standard arguments that lead to positivity bounds.
Therefore, we first review how to parameterize amplitudes in a way such that they are well behaved in the forward limit and also are crossing symmetric.
In particular, we review the Breit parameterization advocated for in \Refc{Grall:2021xxm}.
As we do so, we will clarify and highlight certain aspects of this parameterization that will be particularly relevant in scenarios where multiple fields are present in the EFT.
Subsequently, in \cref{sec:gen_opt_thm} we will derive various bounds from analyticity and the generalized optical theorem using this parameterization.

\subsection{Breit Parameterization}
\label{sec:Breit}

In a Lorentz-invariant theory, $2 \to 2$ scalar scattering amplitudes can be written as a function of just two kinematic parameters, $\amp(s,t)$.
If the UV theory that mediates this interaction is causal,\footnote{If the UV completion is a field theory, then the classic results of Refs.~\cite{Bogoliubov1958,Bremermann:1958zz,Hepp:1964} demonstrate that causality implies analyticity (for a recent review of these points, see \Refc{deRham:2017zjm}). By analyticity, we mean the usual assumption that there exist paths in Hilbert space between degrees of freedom in the IR and UV for which the S-matrix, with external states chosen along this path, is an analytic function of $s$ except at poles and branch cuts.} then for fixed and not arbitrarily large $t$, the amplitude---analytically continued to complex $s$---is an analytic function away from the real axis.
Based on this result, dispersion relations linking IR and UV values of the amplitude can be established.
This gives rise to sum rules and positivity bounds on the Wilson coefficients that specify the IR amplitude, as we discuss in \cref{sec:gen_opt_thm}.

Here we consider theories without full Lorentz invariance.
Specifically, we treat the case where boosts are no longer a good symmetry of the system, but general translations and spatial rotations remain so.
An immediate consequence of broken boost symmetry is that scattering amplitudes are not fully determined by $s$ and $t$ alone, and as established when discussing our conventions, we choose the three additional kinematic variables required to be $\omega_s$, $\omega_t$, and $\omega_u$.
To be completely explicit, as we change the external energy of states for fixed $s$ and $t$, or move between reference frames related by a boost, the scattering amplitude will vary.
The additional variables fundamentally complicate dispersive arguments.
We must now specify how the energies are to be varied as we analytically continue $s$, and a poor choice can introduce obstructions.
For instance, a natural starting point is to compute amplitudes in the center-of-mass (CM) frame, an approach that was explored in \Refc{Baumann:2015nta}.
In the CM frame, however, $\omega_s = \sqrt{s}$, and as the amplitude can explicitly depend on $\omega_s$, we have introduced an additional nonanalyticity that must be accounted for.

An approach that avoids such challenges was proposed in \Refc{Grall:2021xxm}, called the \emph{Breit parameterization}.
To implement the parameterization, regardless of what frame one is working in, the energies are fixed according to\footnote{The form we choose here matches \Refc{Melville:2022ykg}, which differs from \Refc{Grall:2021xxm} by the replacement $\gamma \to \gamma^2$.}
\be
\label{eq:BreitParam}
  \omega_s + \omega_u = 2 \gamma^2 M \qc
  \omega_s - \omega_u = \frac{s-u}{4M},
\ee
which then dictates the individual energies as follows:
\be
\label{eq:omega1-4}
  \omega_1 = \gamma^2 M + \frac{1}{2} \omega_t \qc
  \omega_2 = \frac{s-u}{8M} - \frac{1}{2} \omega_t \qc
  \omega_3 = -\gamma^2 M + \frac{1}{2} \omega_t \qc
  \omega_4 = \frac{u-s}{8M} - \frac{1}{2} \omega_t.
\ee
Through these relations we can trade $\omega_s$ and $\omega_u$ for the parameters $M$ and $\gamma$, such that we can represent any amplitude in terms of $\{s, t, \omega_t, M, \gamma\}$.
The goal is of course to study the analytic properties of forward amplitudes, for reasons that we will review in \cref{sec:gen_opt_thm}.
In the forward limit, where one first takes $\omega_t \to 0$ and then $t \to 0$, the energies are specified by $\omega_{3,4} = -\omega_{1,2}$, and from the Breit parameterization $\omega_1 = \gamma^2 M$ and $\omega_2 = s/4M$.
Importantly, $M$ and $\gamma$ are parameters that are held fixed as we analytically continue in $s$.
As a result, $\omega_1$ is kept constant, whereas $\omega_2$ varies with $s$, but it does so without introducing additional nonanalyticities.

We can also provide a physical interpretation for $\gamma$ and $M$.
To do so, we will primarily restrict our considerations to the forward limit.
First, as the notation suggests, $\gamma^2$ is related to a boost.
To clarify, we introduce $\gamma_s^2 \equiv \omega_s^2/s$ and $\gamma_u^2 \equiv \omega_u^2/u$ as the boosts required to move from the frame in which the $\omega_i$ are defined to the CM frame of the physical $s$ or $u$ channel, respectively.
Using these variables and combining the two relations in \cref{eq:BreitParam}, we have\footnote{The second relation was also given in \Refc{Melville:2022ykg}, and we refer there for further discussion on the physical interpretation of $\gamma$.}
\be
\label{eq:gammadef}
  \gamma^2 = \frac{2(\omega_s^2 - \omega_u^2)}{s-u} = \frac{2(s\gamma_s^2 - u \gamma_u^2)}{s-u}.
\ee
We see that $\gamma^2$ is a weighted combination of the two boosts $\gamma_s$ and $\gamma_u$, chosen in such a manner that it is manifestly $s \leftrightarrow u$ crossing symmetric.
Taking the forward limit, and assuming that the external states satisfy $\omega \gg m$, \cref{eq:gammadef} simplifies to
\be
\label{eq:gfromth}
  \gamma^2 = \frac{1}{\sin^2(\theta/2)} \in [1,\infty).
\ee
Here $\theta$ is the angle between the incident scattering states, $\bk_1$ and $\bk_2$, which will vary between frames, as therefore will $\gamma$.
The minimum value is achieved for $\theta = \pi$, so that the scattering states are back-to-back.
This is the configuration in the CM frame and clarifies why the results of \Refc{Grall:2021xxm} reduce to the earlier relations found in \Refc{Baumann:2015nta} for $\gamma=1$, as the latter worked in the CM frame; we will review these results in \cref{sec:bounds}.
Combined with the above expression for $\gamma$, the first condition in \cref{eq:BreitParam} implies that in the forward limit and for $\omega \gg m$,
\be
  M = \omega_1 \sin^2(\theta/2) \in [0,\omega_1].
\ee
Again the relation between $M$ and $\omega_1$ varies between frames, but for antiparallel scattering states (as in the CM frame), $M = \omega_1$.

\subsection{A Violation of Conventional Analyticity}
\label{sec:acausality}

The Breit parameterization is a convenient formalism for analytically continuing amplitudes in theories with Lorentz violation.
Using \cref{eq:BreitParam}, in the forward limit we have $\omega_s = (s/4M) + \gamma^2 M$, which explicitly depends on $s$.
As $s$-channel propagators involve $\omega_s$, this additional $s$ dependence raises the possibility of an additional pole.
In this section, we show that not only is this possible, but if we consider the case of $\pi \pi \to \pi \pi$ scattering mediated by a second field $\sigma$ where $c_\sigma > c_\pi$, there are frames in which the poles move off the real $s$ axis, in manifest violation of the conventional statements of analyticity, where all nonanalyticities satisfy $\Im(s) = 0$.

While we demonstrate the appearance of  analyticity-violating poles explicitly using the Breit parameterization, a modification of the analytic properties of the S-matrix might also be expected on physical grounds.
In particular, in the conventional proofs of analyticity, causality is invoked by requiring operators commute outside the lightcone.
For theories with Lorentz violation, the external states can have a maximum speed of propagation $c_\mathrm{IR} < 1$ and therefore a contracted causal cone.
One would then immediately worry about the possibility of states propagating with speeds in the range $(c_\mathrm{IR},1]$, as these objects could lead to apparent causality-violating effects from the perspective of the external states, in principle producing nonanalyticities.
Indeed, this concern was discussed in \Refc{Grall:2021xxm}, where for a single field those authors argued that  analyticity in Lorentz-violating theories follows nonperturbatively only when the maximal speed of propagation in the UV completion, $c_\mathrm{UV}$, satisfies $c_\mathrm{UV} \leq c_\mathrm{IR}$. This ensures that there can be no communication outside the IR lightcone between the external, EFT states.
Of course, in a typical dispersive model, one instead expects $c_{\rm UV}$ to be larger than the IR propagation speed. 
Indeed, we will see that in Lorentz-violating theories with multiple degrees of freedom propagating at different speeds, a significant obstruction arises.

Let us perform the explicit calculation for $2\to 2$ scattering of $\pi$ mediated by $\sigma$, with speeds $c_{\pi,\sigma}$.
In order to isolate the additional poles, we will simply consider the $s$-channel contribution as shown below, as the argument proceeds identically for the $u$-channel:
\begin{center} 
  \includegraphics[height=2cm]{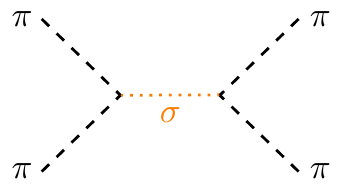} 
\end{center}
The amplitude will involve an $s$-channel propagator (suppressing the $i\epsilon$ for simplicity),
\begin{equation}
  \frac{1}{(\omega_1+\omega_2)^2 - c_\sigma^2 (\vk_1 + \vk_2)^2 - m_\sigma^2}.
\end{equation}
The Mandelstam invariants are set by the external kinematics, and in particular we have $s = \omega_{12}^2 - c_\pi^2 (\bk_1+\bk_2)^2$, so that we can rewrite the propagator as
\begin{equation}
\label{eq:sprop}
  \frac{1}{\omega_s^2 + x (s-\omega_s^2) - m_\sigma^2},
\end{equation}
where we have introduced $x = c_\sigma^2/c_\pi^2$ as the ratio of the speeds.

Let us momentarily set $m_\sigma=0$.
Then the above propagator implies that there is a pole in the S-matrix at $s = \omega_s^2 (1 - 1/x)$.
For Lorentz-invariant theories $c_\sigma = c_\pi = 1$, implying $x=1$, this pole moves to $s=0$, as expected for the exchange of a massless state.
However, for any $x\neq 1$, the location of this pole depends sensitively on how $\omega_s$ is parameterized as we move in the complex $s$ plane in the forward limit. In particular, if $\omega_s$ is complex, the pole can move away from the real $s$ axis, invalidating the standard arguments underlying positivity bounds.
To study the properties of the poles further, we need to adopt a procedure for how $\omega_s$ will vary as we analytically continue $s$ in the forward limit; however, the above discussion identifies a challenge that will appear whenever $c_{\sigma} \neq c_{\pi}$.

To proceed, let us restore $m_{\sigma}$ and adopt the Breit parameterization, which instructs us to take $\omega_s =(s/4M) + \gamma^2 M$.
Doing so, the propagator in Eq.~\eqref{eq:sprop} has two poles, occurring at
\begin{equation}
\label{eq:spm}
  s_\pm = \frac{4M}{x-1} \Bqty{(1-x)\gamma^2 M + 2Mx
  	                           \pm \sqrt{4x M^2 \pqty{x+\gamma^2(1-x)} + m_\sigma^2 (1-x)}}. 
\end{equation}
As expected, an additional pole has emerged.
If we again take $c_\sigma = c_\pi$ (\ie, $x = 1$), then the situation reduces to a single field problem, as far as the kinematics is concerned, and we would expect the additional pole to decouple.
Indeed, taking $x=1+\epsilon$, we find $s_- = m_\sigma^2 + \ord(\epsilon)$, the conventional pole, and $s_+ = 16 M^2/\epsilon - m_{\sigma}^2 - 8 M^2 (\gamma^2-2) + \ord(\epsilon)$, which for sufficiently small $\epsilon$ will lie outside the range of validity of our EFT.

It is also instructive to consider the residues of \cref{eq:sprop} on the two poles; they are given by
\begin{equation}
  \underset{s=s_\pm}{\!\!\!\Res}\amp
    = \mp\frac{2M}{\sqrt{4x M^2 \pqty{x+\gamma^2(1-x)} + m_\sigma^2 (1-x)}}.
\end{equation}
Hence the residue of the conventional pole $s_-$ is indeed positive, although the residue associated with the additional state is negative, indicating it is a ghost.
As this calculation corresponds to the residue of the propagator only, it ignores additional terms in the amplitude coming from the structure of the interaction terms.
However, these are positive for real values of $s$ and thus do not change the conclusions above.

More concerning, however, is that if the quantity under the radical is negative, then both poles occur off the real axis.
The condition for this to occur is
\begin{equation}
\label{eq:complexpoles}
  \frac{x-1}{x} \pqty{ \gamma^2 + \frac{m_\sigma^2}{4M^2 x} } > 1.
\end{equation}
As this condition depends on both $\gamma$ and $M$, the appearance of complex poles will depend on the specific parameters of the scattering, as seen in \cref{sec:Breit}.
Nevertheless, we observe that as the expression in parentheses in \cref{eq:complexpoles} is strictly positive, then if $x \leq 1$ (\ie, $c_\sigma \leq c_\pi$) the left-hand side is strictly negative and the poles are real.
Thus, a necessary---although not sufficient---condition for this process to develop complex poles in its scattering amplitude is $c_\sigma > c_\pi$.

To further simplify the condition, we again consider the case where $m_\sigma$ is parametrically small and can be neglected.
\Cref{eq:complexpoles} then simplifies to $x > \gamma^2/(\gamma^2\,{-}\,1)$.
In the CM frame, where the virtual $\sigma$ is produced at rest, $\gamma = 1$ and the condition cannot be satisfied for any finite $x$.
But otherwise for any given choice of $x > 1$, there exists a $\gamma$ satisfying $x > \gamma^2/(\gamma^2\,{-}\,1)$, or equivalently, $\gamma^2 > x /(x-1)$, such that complex poles appear, obstructing the procedure for obtaining positivity bounds.

The appearance of a second pole with negative residue, as well as the possibility of poles away from the real axis, arose only once we adopted the Breit parameterization.
Yet as both challenges vanish when $x = 1$, the results are suggestive of a more fundamental pathology associated with fast propagating modes mediating ``acausal'' interactions among slower modes, which may persist to other parameterizations.
Since states with negative residue or poles off the real axis invalidate traditional dispersive arguments, for the present work we will always restrict to $x=1$.
Further, we will always consider scenarios in which the single, common speed is parametrically close to unity.
Otherwise, one would anticipate that UV-completing the theories of interest will introduce states with different speeds and immediately regenerate the complex poles.
Nevertheless, the fact that the two-field scenario we consider can be UV-completed into a sensible Lorentz-invariant theory suggests that there is nothing fundamentally sick, and possibly a modified dispersive technique can be developed to handle the issues identified here, thereby allowing our approach to extend beyond $c_{\sigma} = c_{\pi}$.
We do not pursue that possibility here, but note that it would be an interesting and useful direction for future work.

\section{Bounds from the Generalized Optical Theorem on Two-Field Theories}
\label{sec:gen_opt_thm}

Having discussed the Breit parameterization and the conditions under which it leads to forward scattering amplitudes that are analytic in $s$ and crossing symmetric, we are now in a position to employ positivity bounds from dispersion relations.
We will first briefly review the standard arguments leading to positivity bounds in the elastic case and then use that formalism to obtain a stronger set of bounds on inelastic forward amplitudes using the generalized optical theorem.

Traditional positivity bounds on operators in Lorentz-invariant theories arise from performing a contour integral in the complex $s$ plane to extract the relevant Wilson coefficient(s), and then making use of the optical theorem---i.e., unitarity of elastic, forward scattering---to bound the sign.
With the Breit parameterization, we have seen that at forward kinematics the amplitude depends on $s$ alone, $\amp = \amp(s)$.
Therefore the distinction between Lorentz-invariant and -violating theories is immaterial to the consideration of the analytic properties of $\amp(s)$, since $\amp(s)$ is still analytic everywhere other than the real $s$ axis under our criteria mentioned in \cref{sec:acausality}.

By power counting, we will have $\amp(s) \propto s^2$ in the IR for an amplitude generated by quartic dimension-eight contact operators in the EFT.
However, rather than considering a single scalar as in Refs.~\cite{Adams:2006sv,Chandrasekaran:2018qmx,Grall:2021xxm}, or even a superposition of states as in Refs.~\cite{Cheung:2016yqr,Remmen:2019cyz,Remmen:2020vts,Remmen:2020uze,Remmen:2022orj}, we wish to extract the most general bounds possible.
This can be done by making use of the \emph{generalized} optical theorem as in, e.g., Refs.~\cite{Zhang:2020jyn,Trott:2020ebl,Arkani-Hamed:2021ajd,Li:2021lpe}.
For our purposes---as we will eventually be interested in the case of the inflaton and a spectator massless scalar---let us focus on the scattering of two possible states, which for convenience we will label as $\varphi_i$ with flavor index $i = 1, 2$ (\eg, $\varphi_1$ could be the inflaton and $\varphi_2$ a massless spectator).
Then we can define an S-matrix for scattering arbitrary states $\varphi_i(p_1) \varphi_j(p_2) \to \varphi_k(p_3) \varphi_l(p_4)$, with flavor indices $i,j,k,l$, and consider the analytic properties of $\amp_{ijkl}(s)$, which at low energies will scale as $s^2$ times various Wilson coefficients in the two-scalar EFT, which we extract via $M_{ijkl} = \lim_{s\to 0} \partial_s^2 \amp_{ijkl}(s) + {\rm c.c}$.
Given that $M_{ijkl}$ depends only on an energy-squared $s$, it will be symmetric under swapping all incoming and outgoing states.\footnote{This follows from assuming discrete $T$ symmetry, or equivalently, $CP$ symmetry. We assume that the breaking of Lorentz boosts engendered by the inflaton background does not in itself break $CP$.}
Using this fact, combined with Bose symmetry under simultaneously swapping $i\leftrightarrow j$ and $k\leftrightarrow l$, one concludes that a priori there are seven independent components: $M_{1111}$, $M_{2222}$, $M_{1212}$, $M_{1221}$, $M_{1122}$, $M_{1112}$, and $M_{1222}$.
Further, again by Bose symmetry, $\mathcal{A}_{ijkl}(s,t)=\mathcal{A}_{ilkj}(u,t)$, so we have $M_{1221} = \oint {\rm d} s\,\mathcal{A}_{1221}(s,0)/s^3 + {\rm c.c.} = \oint {\rm d}(-s)\,\mathcal{A}_{1122}(-s,0)/(-s)^3 + {\rm c.c.} = M_{1122}$ after relabeling the dummy variable $-s\to s$, leaving a total of six independent terms in $M_{ijkl}$.

Using analytic dispersion relations, we can relate $M_{ijkl}$ to information about the UV and thereby use unitarity to place constraints on the Wilson coefficients.
Though these arguments appear elsewhere in the literature~\cite{Arkani-Hamed:2021ajd,Zhang:2020jyn,Trott:2020ebl}, we will briefly review them here for completeness.
First, we extract $M_{ijkl}$ by performing a contour integral of $\amp_{ijkl}(s)/s^3$ around the origin, then use the analytic structure of the amplitude to deform the contour to one running above and below the entire real $s$ axis:\footnote{The integral over the contour at infinity vanishes, since $\amp_{ijkl}(s)$ scales slower than $s^2$ in the deep UV on general causality grounds~\cite{Arkani-Hamed:2020blm} or alternatively by the Froissart bound (which applies to massive theories and so would require giving $\pi$ and $\sigma$ tiny masses).
In any case, if the momentum scaling of the UV is improved, from a perturbative unitarity standpoint, over the IR EFT scaling, this contour must vanish.
Even though there are massless particles in the theory and thus branch cuts in the amplitude extending to the origin, in a weakly coupled UV completion, such diagrams will be formally subdominant in the couplings and loop expansion and so can be ignored~\cite{Nicolis:2009qm}.}
\bea
  M_{ijkl}
   & = \frac{1}{i\pi} \int_0^\infty \frac{\dd{s}}{s^3}\left[\Disc\amp_{ijkl}(s) + \Disc \amp_{ilkj}(s)\right] + {\rm c.c.} 
   \\& = \frac{2}{\pi} \int_0^\infty \frac{\dd{s}}{s^3} \left[{\rm Im}\amp_{ijkl}(s) + {\rm Im}\amp_{klij}(s) + {\rm Im}\amp_{ilkj}(s)+{\rm Im}\amp_{kjil}(s)\right]\!.
\eea
Here we have used $1 \leftrightarrow 3$ crossing symmetry in the forward limit to set $\amp_{ijkl}(s)=\amp_{ilkj}(-s)$ and Hermitian analyticity~\cite{Stapp:1968zg,Miramontes:1999gd} to equate $\amp_{ijkl}(s) = \amp_{klij}^*(s^*)$.
Now, the conventional optical theorem would, in the elastic case where $(i,j) = (k,l)$, allow us to use unitarity to relate ${\rm Im}\amp(s)$ to the cross section, enforcing positivity of the associated EFT coefficients.
Here, however, let us instead allow arbitrary $i,j,k,l$ and use the generalized optical theorem~\cite{Zhang:2020jyn},
\begin{equation}
2\left({\rm Im}\amp_{ijkl} + {\rm Im}\amp_{klij}\right) = \sum_X \left( \amp_{ij\to X} \amp^*_{kl\to X} + {\rm c.c.}\right)\!.
\end{equation}
Here $\amp_{ij \to X}$ is the amplitude for $\varphi_i(p_1)\varphi_j(p_2) \to X$ with $X$ an arbitrary (possibly multiparticle) intermediate massive state. 
Writing the real and/or imaginary parts of the various $\amp_{ij \to X}$ as some collection of arbitrary real matrices $m^{ij}$, we thus have
\be
  M_{ijkl} = \frac{2}{\pi} \int_0^\infty \frac{\dd{s}}{s^3}
               \sum_m \bqty{m^{ij}(s) m^{kl}(s) + m^{il}(s) m^{kj}(s)},
\ee
where the sum is over all such matrices.
That is, if we use a general index $q$ to represent a sum over the matrix $m$ and an integral over $s$, we can write the bound as~\cite{Arkani-Hamed:2021ajd}:
\be
\label{eq:gen_opt}
  M_{ijkl} = \sum_q \pqty{m_q^{ij} m_q^{kl} + m_q^{il} m_q^{kj}}.
\ee
\Cref{eq:gen_opt} will be the primary result from which we will now derive the consequences of the generalized optical theorem.

Let us define $\vec{u} = m_q^{11}$, $\vec{v} = m_q^{22}$, $\vec{w} = m_q^{12}$, and $\vec{y} = m_q^{21}$.
Then we have the vector relations:
\be
  \begin{split}
    M_{1111} &= 2\sum_q m_q^{11} m_q^{11} = 2\abs{\vec{u}}^2 \\
    M_{2222} &= 2\sum_q m_q^{22} m_q^{22} = 2\abs{\vec{v}}^2 \\
    M_{1221} = M_{1122} &= \sum_q \pqty{m_q^{12} m_q^{21} + m_q^{11} m_q^{22}}
                          = \vec{w}\cdot\vec{y} + \vec{u}\cdot\vec{v} \\
    M_{1212} = M_{2121} &= 2\sum_q m_q^{12} m_q^{12}
                         = 2\abs{\vec{w}}^2 = 2\abs{\vec{y}}^2 \\
    M_{1112} = M_{2111} &= 2\sum_q m_q^{11} m_q^{12}
                         = 2\vec{u}\cdot\vec{w} = 2\vec{u}\cdot\vec{y} \\
    M_{1222} = M_{2221} &= 2\sum_q m_q^{12} m_q^{22}
                         = 2\vec{v}\cdot\vec{w}=2\vec{v}\cdot\vec{y}.
  \end{split}
\ee
In particular, we have the positivity bounds,\footnote{We will characterize bounds derived from the optical theorem as strict inequalities, since cross sections and norms of states are strictly positive.
Of course, as noted in Refs.~\cite{Remmen:2019cyz,Adams:2006sv,Nicolis:2009qm,Chandrasekaran:2018qmx}, if we are restricting EFT amplitudes to a fixed order in couplings or the loop expansion, positivity bounds can soften to weak inequalities, but we will simply write $<$ or $>$ throughout for clarity of notation.}
\be
\label{eq:gtrzero}
  M_{1111} > 0 \qc
  M_{2222} > 0 \qc
  M_{1212} > 0.
\ee
The rest of the bounds implied by unitarity are in terms of three quantities given by the (sums of) vector products $\vec{u}\cdot\vec{v} + \vec{w}\cdot\vec{y}$, $\vec{u}\cdot\vec{w} = \vec{u}\cdot\vec{y}$, and $\vec{v}\cdot\vec{w} = \vec{v}\cdot\vec{y}$.
Let us define the vectors
\be
  \vec{x} = \vec{w}+\vec{y} \qc
  \vec{z} = \vec{w}-\vec{y},
\ee
in terms of which we have
\be
  \begin{split}
    M_{1111} &= 2\abs{\vec{u}}^2 \\
    M_{2222} &= 2\abs{\vec{v}}^2 \\
    M_{1221} &= \frac{1}{4}\abs{\vec{x}}^2 - \frac{1}{4}\abs{\vec{z}}^2
                + \vec{u}\cdot\vec{v} \\
    M_{1212} &= \frac{1}{2}\abs{\vec{x}}^2 + \frac{1}{2}\abs{\vec{z}}^2 \\
    M_{1112} &= \vec{u}\cdot\vec{x} \\
    M_{1222} &= \vec{v}\cdot\vec{x},
  \end{split}
\ee
along with the constraints
\be
  \vec{u} \cdot \vec{z} = \vec{v} \cdot \vec{z} = \vec{x} \cdot \vec{z}=0.
\ee

Without loss of generality, we can take the three vectors $\vec{u},\vec{v},\vec{x}$
to lie in some three-dimensional vector space.
The vector $\vec{z}$ is orthogonal to this space, and it appears in the bounds only through
the unknown positive real parameter $|\vec{z}|^{2}\equiv\mu$.
We can remove it by considering the quantities $M_{1111}, M_{2222}, M_{1112}, M_{1222}$, and
\be
  \begin{split}
    M_{1221}+\frac{1}{4}\mu &= \frac{1}{4}\abs{\vec{x}}^2 + \vec{u}\cdot\vec{v}\\
    M_{1212}-\frac{1}{2}\mu &= \frac{1}{2}\abs{\vec{x}}^2,
  \end{split}
\ee
along with the requirement,
\be
\label{eq:M1212mu}
  M_{1212} > \frac{\mu}{2}.
\ee
We note that the three combinations, $a, b, c$ given by
\be
\label{eq:abc}
  \begin{split}
    a & = \frac{2M_{1221} - M_{1212} + \mu}{\sqrt{M_{1111} M_{2222}}}
        = \frac{\vec{u}\cdot\vec{v}}{\abs{\vec{u}}\abs{\vec{v}}}
        \equiv \cos\theta_a\\
    b & = \frac{M_{1112}}{\sqrt{M_{1111} \pqty{M_{1212}-\frac{1}{2}\mu}}}
        = \frac{\vec{u}\cdot\vec{x}}{\abs{\vec{u}}\abs{\vec{x}}}
        \equiv \cos\theta_b\\
    c & = \frac{M_{1222}}{\sqrt{M_{2222} \pqty{M_{1212}-\frac{1}{2}\mu}}}
        = \frac{\vec{v}\cdot\vec{x}}{\abs{\vec{v}}\abs{\vec{x}}}
        \equiv \cos\theta_c
  \end{split}
\ee
define three angles, $\theta_{a,b,c}$.
Thus, we must have $a,b,c$ between $\pm 1$, \ie,
\be
\label{eq:cos}
  \begin{split}
    \pqty{2M_{1221} - M_{1212} + \mu}^2  < M_{1111} M_{2222} \\
    M_{1112}^2 < M_{1111} \pqty{M_{1212} - \frac{1}{2}\mu}\\
    M_{1222}^2 < M_{2222} \pqty{M_{1212} - \frac{1}{2}\mu}
  \end{split}
\ee
for some unknown $\mu$.

\begin{figure}[t]
\begin{center}
\includegraphics[width=6.5cm]{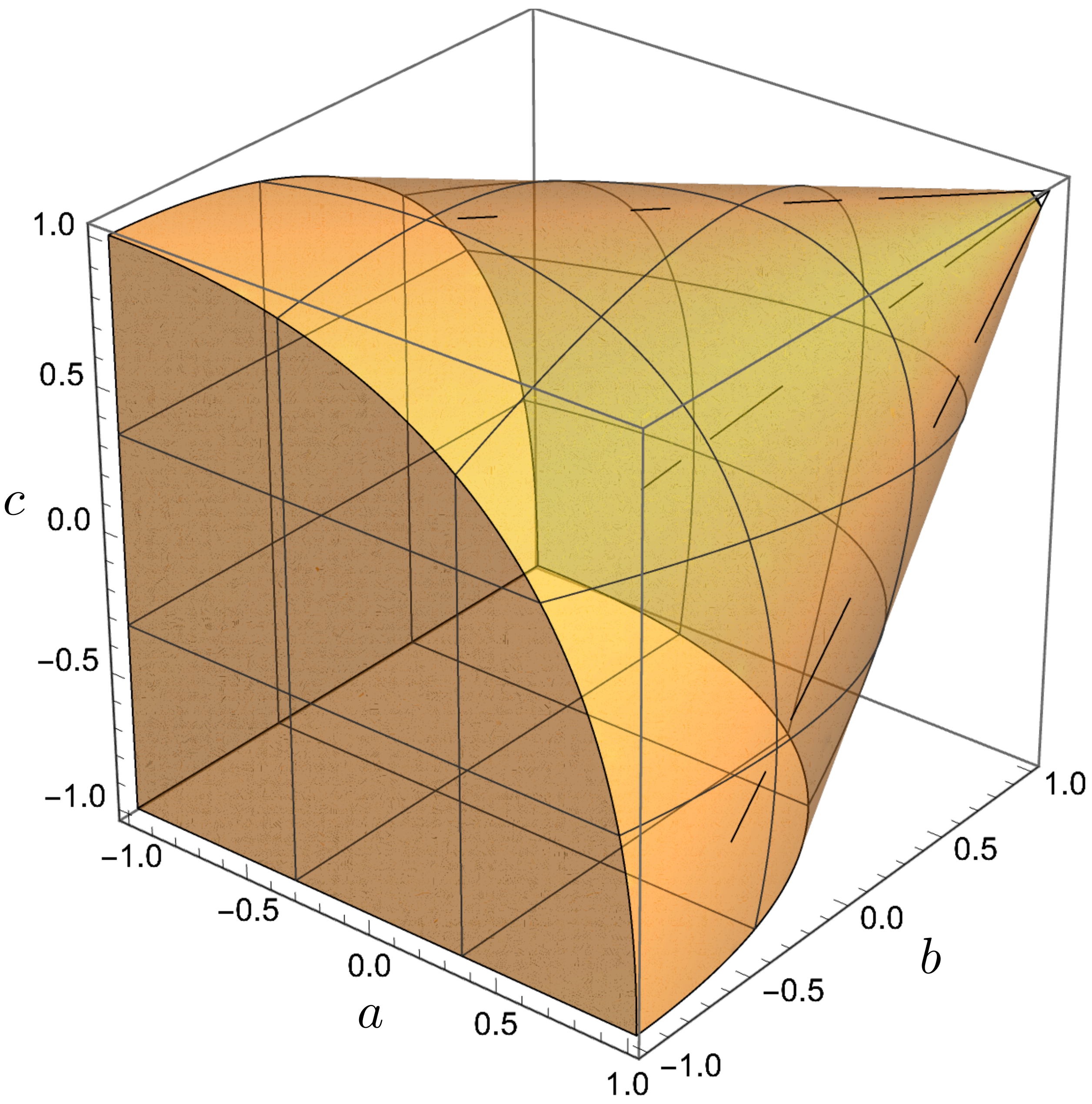}
\end{center}
\vspace{-0.5cm}
\caption{Region in $(a,b,c)$ parameter space, defined in \cref{eq:abc}, permitted by the generalized optical theorem.
}
\label{fig:cone}
\end{figure}

However, to fully characterize the necessary and sufficient bound, we must enforce that $\theta_{a,b,c}$ satisfy triangle inequalities.
That is, let us define a unit-normalized vector $\hat{u}=\vec{u}/|\vec{u}|$ and analogously for $\hat{v}$ and $\hat{x}$.
Then the lengths $\ell_a = |\hat{u}-\hat{v}|$, $\ell_b = |\hat{u}-\hat{x}|$, and $\ell_c = |\hat{v}-\hat{x}|$ each form the base of a different isosceles triangle, in each of which the
two equal sides are of unit length.
The angle of the vertex subtending the base of a given triangle is $\theta_a$, $\theta_b$, or $\theta_c$, respectively, which satisfy $\ell_{a}=2\sin(\theta_{a}/2)=\sqrt{2(1-a)}$ and similarly for $b$
and $c$.
The triangle inequality requires $\ell_a < \ell_b + \ell_c$, $\ell_b < \ell_c + \ell_a$,
and $\ell_a < \ell_b + \ell_c$.
Hence,
\be
  \begin{split}
    \sqrt{1-a} &< \sqrt{1-b}+\sqrt{1-c}, \\
    \sqrt{1-b} &< \sqrt{1-c}+\sqrt{1-a}, \\
    \sqrt{1-c} &< \sqrt{1-a}+\sqrt{1-b}.
  \end{split}
\ee
This is equivalent to the single requirement,
\be
  4(1+ab+bc+ca) > (1+a+b+c)^2,
\ee
depicted in \cref{fig:cone}.
The remaining bound is thus
\begin{align}
    1 + &\frac{(2M_{1221} - M_{1212} + \mu)
 	           \pqty{M_{1112} \sqrt{M_{2222}} + M_{1222} \sqrt{M_{1111}}}
 	           \sqrt{M_{1212}-\frac{1}{2}\mu}}
              {M_{1111} M_{2222} \pqty{M_{1212}-\frac{1}{2}\mu}} \nonumber \\
        \label{eq:triangle}
        &+ \frac{M_{1112} M_{1222} \sqrt{M_{1111} M_{2222}}}
                {M_{1111} M_{2222} \pqty{M_{1212}-\frac{1}{2}\mu}} \\
        & > \frac{1}{4}
                 \pqty{ 1 + \frac{2M_{1221} - M_{1212} + \mu}{\sqrt{M_{1111} M_{2222}}}
     	                + \frac{M_{1112}}{\sqrt{M_{1111}\pqty{M_{1212}-\frac{1}{2}\mu}}}
     	                + \frac{M_{1222}}{\sqrt{M_{2222}\pqty{M_{1212}-\frac{1}{2}\mu}}} }^2. \nonumber
\end{align}
Together, the existence of some $\mu > 0$ for which \cref{eq:gtrzero,eq:M1212mu,eq:cos,eq:triangle} are satisfied comprises the necessary and sufficient condition for the four-point scattering of the two massless scalars to comply with unitarity and analyticity.

We can immediately see that the bounds from generalized unitarity are stronger than those obtainable from elastic scattering (\eg, of definite superpositions of $\phi_i$ states) alone.
For example, the presence of $\mu$, a UV quantity invisible to the IR EFT, in the bounds is indicative of the qualitative difference made by the generalized optical theorem over traditional dispersive positivity bounds.
If $\mu$ takes its maximal value of $2M_{1212}$, a consequence is that the flavor-violating Wilson coefficients $M_{1112}$ and $M_{1222}$ vanish, and the bounds reduce to the statement that $4(M_{1221} + M_{1212})^2 < M_{1111} M_{2222}$.
On the other hand, if $\mu = 0$, then the flavor-violating terms have their weakest upper bound, \eg, $M_{1112} < \sqrt{M_{1111} M_{1212}}$.
Comparing the bounds in \cref{eq:cos} with \cref{eq:triangle}, it can be checked that the latter implies a stronger bound.
However, if there is a $\phi_i \to -\phi_i$ symmetry that implies $M_{1222} = M_{1112} = 0$, and correspondingly, \cref{eq:triangle} does not give any new constraint over \cref{eq:cos}.

What is the physical significance of $\mu$?
It is straightforward to see that the $\mu$ parameter encodes the ($s^{-3}$-weighted) spectral density of odd-spin states in the UV.
By definition, $\mu = \sum_q (m_q^{12} - m_q^{21})^2$, so nonzero $\mu$ corresponds to some UV state(s) $X$ coupling to $\phi_1(k)\phi_2(p) - \phi_2(k)\phi_1(p)$ for some momenta $p$ and $k$, and by Bose symmetry the coupling to $X$ must go as an odd power of $(p-k)$, which by Lorentz invariance requires $X$ to possess an odd number of spacetime indices.

\section{Inflationary Effective Field Theories}
\label{sec:infeft}

\begin{figure}[t]
\begin{center}
\includegraphics[width=0.7\textwidth]{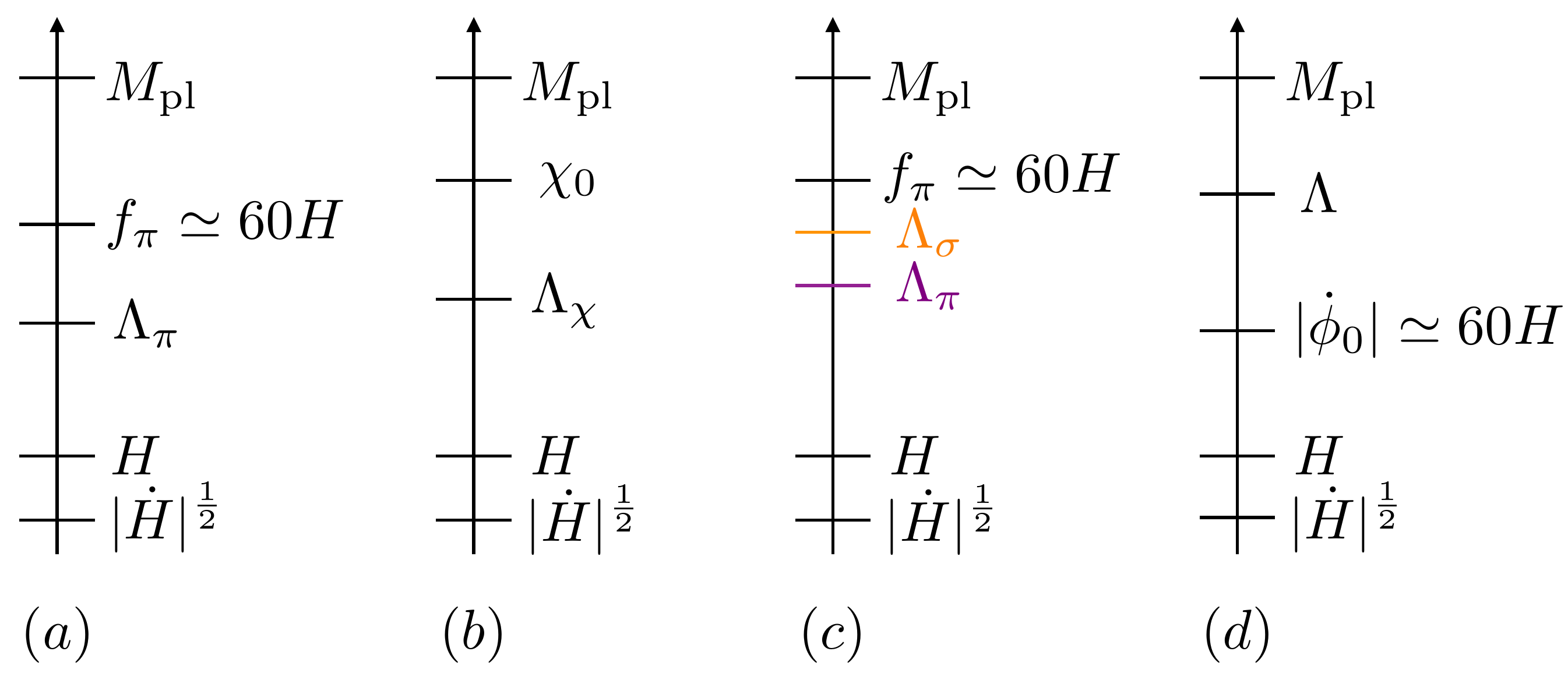}
\end{center}
\vspace{-0.5cm}
\caption{Hierarchy of scales for various EFTs.
From left to right: (a) Goldstone EFT (\cref{sec:GoldstoneEFT}), (b) EFT of multifield inflation with adiabatic perturbations (\cref{sec:mult_EFT_adia}), (c) EFT of multifield inflation with both adiabatic and isocurvature perturbations (\cref{sec:iso_multifield}), (d) Lorentz-invariant EFT of multifield inflation in the slow-roll regime (\cref{sec:mult_field_sr}).
See text for the definitions of the various parameters.
}
\label{fig:scales}
\end{figure}

We now turn to the inflationary EFTs that we wish to constrain using the techniques from the previous sections.
For each EFT, we will write down all the leading operators up to mass dimension eight.
In \cref{sec:bounds} we will discuss the analyticity and unitarity bounds on these EFTs.
We summarize the various scales in these EFTs in \cref{fig:scales} and define them in more detail in the respective subsections.

\subsection{Theories with Adiabatic Perturbations}

We begin our discussion by considering EFTs that describe purely adiabatic fluctuations; subsequently, we will extend the discussion to include isocurvature fluctuations.

\subsubsection{Goldstone EFT of Single-field Inflation}
\label{sec:GoldstoneEFT}

We start with a review of the classic Goldstone EFT of inflation, following Ref.~\cite{Cheung:2007st}.
Here our primary goal is to establish the notation that we use in the subsequent discussions.

In this EFT description, the inflaton perturbations are treated as Goldstone fluctuations arising from spontaneous breaking of time translation symmetry.
To derive the effective action for the inflaton perturbations we can start on a fixed time slice, the ``unitary gauge,'' and write all the terms that are consistent with the unbroken three-dimensional spatial diffeomorphisms.
These terms can be constructed from quantities such as $\{g^{00},\, g^{0\mu} A_\mu,\, K_{\mu\nu},\, t\}$.
Here $A_\mu$ is any vector and $K_{\mu\nu}$ is the extrinsic curvature of the chosen time slice.
To restore the inflaton perturbation $\pi(x)$, we can perform a time diffeomorphism $t \to t + \pi(x)$, which moves us away from the unitarity gauge fixed time slice.
Since terms describing the curvature of the hypersurface, such as $K_{\mu\nu}$, contain extra derivatives, they are typically subdominant and we do not consider them in the following.\footnote{As an exception, in the limit where the speed of inflaton fluctuations becomes very small, such higher derivative terms can become important~\cite{Arkani-Hamed:2003juy}.}
Therefore at leading order in derivatives, we can focus on terms made up of $g^{00}, g^{0\mu} A_\mu, t$. The time diffeomorphism acts on these quantities as
\be
\label{eq:time_diff}
  \begin{split}
    f(t) &\to f(t) + \dot{f}(t) \pi(x), \\
    g^{00} &\to \pqty{\delta^0_\mu + \partial_\mu\pi}
                  \pqty{\delta^0_\nu + \partial_\nu\pi} g^{\mu\nu}
                = g^{00} + 2g^{0\mu} \partial_\mu\pi
                  + g^{\mu\nu} \partial_\mu\pi \partial_\nu\pi, \\
    g^{0\mu} A_\mu &\to (\delta^0_\nu + \partial_\nu\pi) g^{\mu\nu} A_\mu
                        = A^0 +  A^\mu\partial_\mu\pi.
\end{split}
\ee
Since we will be interested in energies above $|\dot{H}|^{1/2}$, we can ignore mixing between $\pi$ and metric perturbations $\delta g^{\mu\nu}$.
Then we can just use the background metric $\bar{g}_{\mu\nu}$ to reduce the right-hand side of \cref{eq:time_diff} to
\begin{equation}
\label{eq:time_diff2}
  \begin{split}
    f(t)& \to f(t) + \dot{f}(t) \pi(x), \\
    g^{00} &\to -1-2\dot{\pi} + \bar{g}^{\mu\nu} \partial_\mu\pi \partial_\nu\pi, \\
    g^{0\mu}A_\mu &\to (\delta^0_\nu + \partial_\nu\pi) \bar g^{\mu\nu} A_\mu
                        = A^0 +  A^\mu\partial_\mu\pi.
  \end{split}
\end{equation}

If we specialize to the case of single-field inflation, there is no vector field like $A^\mu$ contributing to the action.
Therefore the effective Lagrangian for inflaton fluctuations is given purely as an expansion in $(g^{00}+1)^n$.\footnote{We will also work in the limit where the explicit time dependence of an EFT coefficient $c(t)$ is suppressed, i.e., $|\dot{c}(t)|\ll H c(t)$.}
Introducing $\pi$ with the above time diffeomorphism, we arrive at the adiabatic single-field (AS) Lagrangian,
\begin{equation}
  \mathcal{L}_\mathrm{AS}
    = \Mpl^2\dot{H}(\partial_\mu\pi)^2
      + \sum_{n=2}^\infty \frac{M_n^4}{n!} \left[-2\dpi+(\partial_\mu\pi)^2\right]^n,
\end{equation}
where we have introduced the notation $(\partial_\mu\pi)^2 = \bar{g}^{\mu\nu} \partial_\mu\pi \partial_\nu\pi$, and where $M_n^4$ are arbitrary couplings.
The kinetic term arises from a term linear in $g^{00}$ and can be fixed by demanding an expansion around the correct de Sitter (dS) background~\cite{Cheung:2007st}.
Keeping up to fourth order in fluctuations,
\begin{equation}
\label{eq:LAS_1}
  \begin{split}
    \mathcal{L}_\mathrm{AS}
     & = \Mpl^2\dot{H}(\partial_\mu\pi)^2
          + M_2^4 \left[2\dpi^2 - 2\dpi(\partial_\mu\pi)^2
       	  + \frac{1}{2} (\partial_\mu\pi)^2 (\partial_\nu\pi)^2\right] \\
     &\qquad - M_3^4 \left[\frac{4}{3} \dpi^3 - 2\dpi^2(\partial_\mu\pi)^2\right]
               + \frac{2}{3} M_4^4 \dpi^4.
  \end{split}
\end{equation}
It is now manifest that unless $M_2 = 0$, the speed of inflaton propagation $\cs$ will differ from 1, since there is no symmetry protecting that value.
We can parameterize the deviation as
\be
  \cs^2 = \frac{\Mpl^2\dot{H}}{\Mpl^2\dot{H} - 2M_2^4}.    
\ee
Using this parameterization we can define a canonically normalized inflaton mode, $\pi_{\rm c} \equiv f_\pi^2\pi$ with $f_\pi^2 = (2\Mpl^2|\dot{H}|\cs)^{1/2}$.
Written in terms of canonically normalized fields, the AS Lagrangian becomes
\begin{equation}
\label{eq:goldstone_eft}
  \begin{split}
  \mathcal{L}_\mathrm{AS}
    &= \frac{1}{\cs^3} \left[\frac{1}{2}\dpi_c^2 - \cs^2\frac{1}{2a^2}(\delipi_c)^2\right] \\
    &\qquad - \frac{1}{4\cs f_\pi^2}\pqty{1-\frac{1}{\cs^2}}
                \left[-2\dpi_c(\partial_\mu\pi_{\rm c})^2
            	      + \frac{1}{2f_\pi^2}(\partial_\mu\pi_\mathrm{c})^2
            	                            (\partial_\nu\pi_\mathrm{c})^2\right] \\
    &\qquad  - \frac{4}{3} \frac{M_3^4}{f_\pi^6} \dpi_c^3
               + \frac{2M_3^4}{f_\pi^8} \dpi_c^2(\partial_\mu\pi_{\rm c})^2
               + \frac{2}{3}\frac{M_4^4}{f_\pi^8}\dpi_c^4.
  \end{split}
\end{equation}
We note that even in the absence of $M_{n>2}$ operators, a sound speed $\cs<1$ induces a dimension-six operator and gives rise to a strong coupling scale~\cite{Cheung:2007st},
\be
  \Lambda_\pi^4 \simeq 8\pi^2 f_\pi^4 \frac{\cs^4}{1-\cs^2}.
\ee
Noting the fact that the magnitude of the scalar power spectrum fixes $f_\pi \simeq 60H$, a scenario with $\cs\ll 1$ corresponds to a strong coupling scale parametrically close to $H$ itself.
For $\cs$ parametrically close to $1$, which will be our focus in the following discussion, we have $\Lambda_\pi \gg f_\pi$, and the EFT cutoff scale is instead determined by the operators with coefficients $M_3$ and $M_4$ in \cref{eq:goldstone_eft}.
We summarize this hierarchy of scales in \cref{fig:scales}~(a) with $\Lambda_\pi$ determining the EFT cutoff above which the EFT needs to be UV completed.
The relation $\Lambda_\pi > H$ ensures that we can compute scattering amplitudes in a subhorizon regime while $\Lambda_\pi < f_\pi$ ensures the validity of the Goldstone description.
The hierarchy $|\dot{H}|\ll H^2$ is obeyed given our assumption of a quasi-dS background around which we are expanding the perturbations.

It will be convenient to rewrite $\mathcal{L}_\mathrm{AS}$ once more, which we do by introducing dimensionless coefficients $c_i$ defined via $M_n^4=c_n f_\pi^4/\cs^{2n-1}$ and rescaling $\tilde{\vec{x}} = \vec{x}/\cs$~\cite{Baumann:2015nta}, after which we have
\begin{multline}
\label{eq:LAS_final}
  \cs^3\mathcal{L}_\mathrm{AS}
    = -\frac{1}{2}(\tilde{\partial}_\mu\pi_{\rm c})^2
      + \frac{1}{\cs^2f_\pi^2}\left[-2c_2 \dpic(\tilde{\partial}_\mu\pi_{\rm c})^2 - \left(8c_2^2 + \frac{4}{3}c_3\right)\dpic^3\right] \\ 
      + \frac{1}{\cs^4f_\pi^4}\left[\left(8c_2^3 + 8c_2c_3 + \dfrac{2}{3}c_4\right)\dpic^4 + \left(2c_3+4c_2^2\right) \dpic^2 (\tilde{\partial}_\mu\pi_{\rm c})^2 + \frac{1}{2}c_2 (\tilde{\partial}_\mu\pi_{\rm c})^4 \right]\!,
\end{multline}
where we have grouped the interactions into mass dimension six and eight.
Here $c_2=(1-\cs^2)/4$, and $\tilde{\partial}_\mu$ is defined with respect to $\{t,\tilde{\vec{x}}\}$.
This variable change has restored a fake Lorentz invariance into the Lagrangian that is convenient for studies of analytic properties of the amplitudes.
Although we have kept $\cs$ explicit in all expressions above, we emphasize once more that at present the nonrelativistic dispersive tools we will deploy demand we focus on scenarios where $\cs$ is parametrically close to 1, and therefore $\mathcal{L}_{\rm AS}$ will simplify accordingly.

\subsubsection{EFT of Multifield Inflation with Adiabatic Perturbations}\label{sec:mult_EFT_adia}

Next we consider a scenario where the adiabatic fluctuations are produced by a field (or fields) that are different from those responsible for the background inflationary expansion.
Examples include the case of multifield inflation (for reviews see~\cite{Bassett:2005xm, Malik:2008im}) as well as the scenario of the curvaton~\cite{Linde:1996gt,Enqvist:2001zp,Lyth:2001nq,Moroi:2001ct}, whose fluctuations are converted to adiabatic perturbations when it decays around or after the end of inflation.

In a multifield scenario, the (comoving) curvature perturbation $\zeta$ can receive contributions from multiple light fields~~\cite{Bassett:2005xm, Malik:2008im} which we denote by $\chi_i$.
To ensure that these fields generate superhorizon curvature perturbations, they must all be light.
A natural way to ensure this is through symmetry.
For example, in \Refc{Senatore:2010wk} such EFTs were constructed by protecting the lightness of the $\chi_i$ fields with a shift symmetry or an approximate supersymmetry.
The full EFT involving all $\chi_i$ can contain various terms, depending on the governing symmetry.
However, to describe the leading NG signature from such shift-symmetric, adiabatic multifield (AM) EFTs we can focus on a simplified parameterization along the lines of \Refc{Smith:2015uia}, 
\begin{equation}\begin{aligned}
\mathcal{L}_{\rm AM} &= -\frac{1}{2}(\partial_\mu\chi)^2 + \frac{e_1}{\Lambda_\chi^4}\dot{\chi}^4 + \frac{e_2}{\Lambda_\chi^4}\dot{\chi}^2(\partial_i\chi)^2 + \frac{e_3}{\Lambda_\chi^4}(\partial_i\chi)^4,\\
&= -\frac{1}{2}(\partial_\mu\chi)^2 + \frac{e_1+e_2+e_3}{\Lambda_\chi^4}\dot{\chi}^4 + \frac{e_2+2e_3}{\Lambda_\chi^4}\dot{\chi}^2(\partial_\mu\chi)^2 + \frac{e_3}{\Lambda_\chi^4}(\partial_\mu\chi)^4.
\label{eq:LAM}
\end{aligned}\end{equation}
Here $e_1,e_2,e_3$ are EFT coefficients, and we have also imposed a $\ZZ_2$ symmetry on $\chi$.
The scale $\Lambda_\chi$ determines the cutoff scale for the EFT.
With this additional symmetry we will be able to connect the positivity bounds that we derive to observational constraints on the primordial trispectrum derived from Planck data~\cite{Planck:2019kim}, which was obtained using the EFT in \cref{eq:LAM}.
To be more general, however, we will relax the assumption of the $\ZZ_2$ symmetry in \cref{sec:iso_multifield} when we consider the presence of isocurvature perturbations.

As an example of the above EFT, we can consider the curvaton scenario~\cite{Linde:1996gt,Enqvist:2001zp,Lyth:2001nq,Moroi:2001ct}.
We imagine the perturbations of the inflaton field to be very small during inflation.
On the other hand, the $\chi$ field, while subdominant in terms of homogeneous energy density, acquires isocurvature perturbations of order $H/(\pi \chi_0)\sim 10^{-5}$.
Here $\chi_0$ is a typical ``misaligned'' value of the homogeneous $\chi$ field during inflation.
After the end of inflation, the $\chi$ field dilutes like matter and eventually dominates the energy density of the universe. 
It subsequently decays into standard model (SM) particles.
As a result, the SM bath inherits perturbations of the $\chi$ field, and thus isocurvature perturbations are converted into adiabatic perturbations.

We summarize the various scales involved in \cref{fig:scales}~(b).
The assumption of quasi-dS spacetime still implies $|\dot{H}|\ll H^2$.
The EFT cutoff $\Lambda_\chi$ controls the interaction of the curvaton fluctuations and can be parametrically small compared to $\chi_0$ without violating EFT power counting.

\subsection{Theories with Isocurvature Perturbations}

Above we considered EFTs that described scenarios where the late-time perturbations were purely adiabatic.
We now extend the discussion to theories where in addition to the adiabatic perturbations, isocurvature perturbations are also generated by a second light field $\sigma$, and these isocurvature perturbations leave imprints on the cosmic microwave background (CMB).
Similar to $\chi$, we will consider $\sigma$ to be a spectator field during inflation, and it will therefore possess a subdominant energy density compared to the inflationary background.
However, whereas $\chi$ could decay into SM states, and thereby erase the isocurvature fluctuations, we imagine that $\sigma$ persists and thereby gives rise to isocurvature perturbations at late times.
A classic example is when $\sigma$ is an axionlike particle and constitutes the dark matter (DM).
The isocurvature fluctuations in $\sigma$ would then manifest as DM isocurvature~\cite{Marsh:2015xka}, which is subject to various CMB constraints~\cite{Planck:2018jri}.
Similarly to the discussion above, we will assume that there is a shift symmetry protecting the mass of $\sigma$, and therefore we expect it to couple derivatively to itself and to the adiabatic fluctuations.

\subsubsection{EFT of Multifield Inflation with Isocurvature Perturbations}\label{sec:iso_multifield}

To describe isocurvature perturbations, we begin with a scenario that contains two dynamical degrees of freedom: the inflaton fluctuations and a spectator scalar $\sigma$, which sources isocurvature perturbations.
While we refer to this scenario as ``multifield inflation,'' we reemphasize that it is only the inflaton that drives the homogeneous expansion, for which $\sigma$ does not play any major role.
We will remain agnostic as to the origin of this homogenous expansion, and therefore model the inflaton fluctuations by the Goldstone degree of freedom $\pi$ exactly as in the Goldstone EFT of inflation reviewed in \cref{sec:GoldstoneEFT}.
An EFT containing both $\pi$ and $\sigma$ can then be constructed in a fashion similar to the treatment above.
However, we now have two additional components from which we can construct interactions, $g^{0\mu}\partial_\mu\sigma$ and $(\partial_\mu\sigma)^2$, in addition to $\delta g^{00}\equiv g^{00}+1$, which we used previously.
Using these objects, we can schematically write down all possible operators up to mass dimension eight in the isocurvature multifield (IM) EFT,\footnote{We will ignore terms involving extrinsic curvature and work at the leading order in the derivative expansion as before, both for simplicity and as such terms are expected to be subdominant.}
\begin{align}\label{eq:L_IM}
\mathcal{L}_{\rm IM} = \mathcal{L}_{\rm AS} + \mathcal{L}_{1 \sigma} +  \mathcal{L}_{2 \sigma} +  \mathcal{L}_{3 \sigma} +  \mathcal{L}_{4 \sigma}.
\end{align}
Here we have organized the various terms as an expansion in the number of $\sigma$ fluctuations present, and we have further restricted ourselves to only work up to $\mathcal{L}_{4 \sigma}$, as our focus for positivity will be on tree-level $2 \to 2$ scattering, making higher-point interactions irrelevant.
As before, we will work in the limit where the speeds of fluctuations for both $\pi$ and $\sigma$ are parametrically close to unity, given the discussion in \cref{sec:twofield}.
Therefore, our starting point is \cref{eq:LAS_final} with $\cs\simeq 1$,
\begin{multline}
\label{eq:LAS_cs1}
    \mathcal{L}_{\rm AS}= -\frac{1}{2}(\partial_\mu\pi_{\rm c})^2    + \frac{1}{f_\pi^2}\left[-2c_2 \dpic(\partial_\mu\pi_{\rm c})^2 - \left(8c_2^2 + \frac{4}{3}c_3\right)\dpic^3\right]\\ 
    + \frac{1}{f_\pi^4}\left[\left(8c_2^3 + 8c_2c_3 + \dfrac{2}{3}c_4\right)\dpic^4 + \left(2c_3+4c_2^2\right) \dpic^2 (\partial_\mu\pi_{\rm c})^2 + \frac{1}{2}c_2 (\partial_\mu\pi_{\rm c})^4 \right]\!.
\end{multline}
Note that taking the speed parametrically close to unity implies that we must have $c_2\ll 1$ given $c_2 = (1-\cs^2)/4$.

To include the isocurvature fluctuations, we must consider the full set of terms generated from the additional building blocks $g^{0\mu}\partial_\mu\sigma$ and $(\partial_\mu \sigma)^2$, working order by order in $\sigma$.
Up to quartic order in the interactions, these terms are:
\bea
\mathcal{L}_{1\sigma} &=  d_1 f_\pi^2(\delg)(g^{0\mu}\partial_\mu\sigma)+  d_2f_\pi^2(\delg)^2(g^{0\mu}\partial_\mu\sigma)+ d_3f_\pi^2(\delg)^3(g^{0\mu}\partial_\mu\sigma) \\
\mathcal{L}_{2\sigma} &=
d_4(g^{0\mu}\partial_\mu\sigma)^2 +
d_5(\delg)(g^{0\mu}\partial_\mu\sigma)^2 + d_6(\delg)^2(g^{0\mu}\partial_\mu\sigma)^2 \\&\qquad + 
d_7 (\delg) (\partial\sigma)^2 + 
d_8 (\delg)^2 (\partial\sigma)^2 \\
\mathcal{L}_{3\sigma} &=
\frac{d_9}{f_\pi^2}(g^{0\mu}\partial_\mu\sigma)^3 +
\frac{d_{10}}{f_\pi^2}(g^{0\mu}\partial_\mu\sigma)(\partial\sigma)^2 +
\frac{d_{11}}{f_\pi^2}(\delg)(g^{0\mu}\partial_\mu\sigma)^3 \\& \qquad +
\frac{d_{12}}{f_\pi^2}(\delg)(g^{0\mu}\partial_\mu\sigma)(\partial\sigma)^2 \\
\mathcal{L}_{4\sigma} &= \frac{d_{13}}{f_\pi^4}(g^{0\mu}\partial_\mu\sigma)^4 +
\frac{d_{14}}{f_\pi^4}(g^{0\mu}\partial_\mu\sigma)^2(\partial\sigma)^2 +
\frac{d_{15}}{f_\pi^4}(\partial\sigma)^4.
\label{eq:IM_first} 
\eea
Here we have normalized all the operators with $f_\pi=(2\Mpl^2|\dot{H}|\cs)^{1/4}\simeq (2\Mpl^2|\dot{H}|)^{1/4} \simeq 60 H$,  the mass scale we specified previously, and have defined dimensionless coefficients $d_i$.
To reveal the interactions between $\pi$ and $\sigma$, we transform $g^{0\mu}\partial_\mu\sigma$ to unitary gauge,
\begin{align}
g^{0\mu}\partial_\mu\sigma \to -(1+\dot{\pi})\dot{\sigma} + \frac{1}{a^2}\partial_i\pi\partial_i\sigma.
\end{align}
Furthermore, since our goal is to compute various $2\to 2$ amplitudes, it is convenient to arrange terms in $\mathcal{L}_{\rm IM}$ by the order of the interactions, $\mathcal{L}_{\rm IM} = \mathcal{L}_{\rm IM, quadratic} + \mathcal{L}_{\rm IM, cubic} + \mathcal{L}_{\rm IM, quartic}$.
At the quadratic level we have\footnote{As we are working for $\cs \simeq 1$, almost everywhere we take $\cs = 1$, except in places where it would cause an operator coefficient to identically vanish, as is the case for $c_2$. In other words, here we consider scenarios with nonzero but small $c_2\ll 1$, ensuring $\cs\simeq 1$.}
\bea
\mathcal{L}_{\rm IM, quadratic} &= \Mpl^2\dot{H}(\partial_\mu\pi)^2 -\frac{1}{2} (\partial_\mu\sigma)^2 + 2 M_2^4\dot{\pi}^2 + 2d_1f_\pi^2\dot{\pi}\dot{\sigma} + d_4\dot{\sigma}^2\\
&\simeq -\frac{1}{2}(\partial_\mu\pi_{\rm c})^2 -\frac{1}{2} (\partial_\mu\sigma)^2 + 2 c_2\dot{\pi}_{\rm c}^2 + 2d_1\dot{\pi}_{\rm c}\dot{\sigma} + d_4\dot{\sigma}^2.
\eea
Here $\pi_{\rm c} = f_\pi^2 \pi$ is the canonically normalized inflaton fluctuation as before.
In order to diagonalize the kinetic term, if we assume that $c_2,d_1,d_4$ are EFT coefficients independent of the fields themselves, we can perform an orthogonal rotation, $\tilde{\pi}_c = \pi_{\rm c}\cos\alpha+\sigma\sin\alpha$ and $\tilde{\sigma}=-\pi_{\rm c}\sin\alpha+\sigma\cos\alpha$.
Moving to the rotated basis, we have
\be
\mathcal{L}_{\rm IM, quadratic} \simeq -\frac{1}{2}(\partial_\mu\tilde{\pi}_c)^2 -\frac{1}{2} (\partial_\mu\tilde{\sigma})^2 + d_\pi \dot{\tilde{\pi}}_c^2 + d_\sigma \dot{\tilde{\sigma}}^2.
\ee
where
\be
d_{\pi,\sigma} = \frac{1}{2} \left(2 c_2 + d_4 \mp \sqrt{4 c_2^2 + 4 d_1^2 - 4 c_2 d_4 + d_4^2}\right)
\ee
with $-(+)$ sign corresponding to $d_{\pi(\sigma)}$. 
The rotation angle can be determined explicitly as
\be
\tan\alpha = \dfrac{2 d_1}{2 c_2 - d_4 - \sqrt{4 d_1^2 + d_4^2 - 4 d_4 c_2 + 4 c_2^2}}. 
\ee

We next summarize the cubic and quartic interaction terms following from \cref{eq:IM_first} in \cref{tab:cubic_diff_basis,tab:quartic_diff_basis}, respectively.
To clarify the form of the terms appearing, the tables are organized so that each row contains the vertices with the same number of $\sigma$ fields.
Further, for simplicity we have denoted $(\partial\pi_{\rm c})^2 \equiv (\partial_\mu\pi_{\rm c})^2$, $(\partial\pi_{\rm c})^4 \equiv (\partial_\mu\pi_{\rm c})^2 (\partial_\nu\pi_{\rm c})^2$ and analogously for operators involving only $\sigma$, along with the notation $(\partial\pi_{\rm c}\cdot\partial\sigma) \equiv \partial_\mu\pi_{\rm c} \partial^\mu\sigma$.

Here, for simplicity, we have suppressed the higher-dimension interactions of {\it both} $\pi_{\rm c}$ and $\sigma$ by the same scale $f_\pi$.
While for $\mathcal{O}(1)$ values of the coefficients $c_i$ we expect the $\pi_{\rm c}$ sector to have an EFT cutoff close to $f_\pi$, in principle, the $\sigma$ sector could have an EFT cutoff parametrically different than $f_\pi$.
Nonetheless, such a difference can still be captured by the appropriate coefficients $d_i$, except that those coefficients then would not be $\mathcal{O}(1)$.
We also note that the interactions between $\pi_{\rm c}$ and $\sigma$ can also naturally give rise to similar EFT cutoff scales.
As an example, in the presence of a $\dpic^2\dsig/f_\pi^2$ interaction, an interaction of the type $\dsig^3/(16\pi^2 f_\pi^2)$ would be generated at one loop with a similar suppression scale $\sim 4\pi f_\pi$.
We summarize the situation in \cref{fig:scales}~(c), where we have denoted the EFT scales for $\pi_{\rm c}$ and $\sigma$ sectors as $\Lambda_\pi$ and $\Lambda_\sigma$, respectively.
For $\mathcal{O}(1)$ values of $c_i,d_i$, we expect $\Lambda_\pi\sim \Lambda_\sigma \sim f_\pi$.

All terms in \cref{tab:cubic_diff_basis,tab:quartic_diff_basis} are derived in the unrotated basis.
However, it can be checked that a field rotation will only change the coefficients of various terms but not generate any new operator.
Therefore, with a slight abuse of notation, we will continue to use the same coefficient as in the tables but assume a field rotation has been performed such that there is no kinetic mixing term between $\pi$ and $\sigma$.
We further note that in parts of the following discussion, we will also focus on a case where there is a $\ZZ_2$ symmetry acting on $\sigma$.
This symmetry, along with reducing the parameter space in a controlled way, will also remove the kinetic mixing.

\renewcommand{\arraystretch}{1.1}
\begin{table}[!htb]
    \begin{center}
    \begin{tabular}{|c|c|}
    \hline
    	$\pidc\delpisqc$ & $\pidc^3$ \\
    	\hline
    	{\color{red}$-2M_2^4/f_\pi^6$} & {\color{blue}$-\frac{4}{3}M_3^4/f_\pi^6$}\\
    	\hline
    \end{tabular}
    
    \begin{tabular}{|c|c|c|}
    \hline
    	$\delpisqc\sigd$ & $\pidc^2\sigd$ & $\pidc(\delpic\cdot\delsigma)$ \\
    	\hline
    	{\color{brown}$-d_1/f_\pi^2$} & {\color{cyan}$-4d_2/ f_\pi^2$} & {\color{brown}$-2d_1/ f_\pi^2$}\\
    	\hline
    \end{tabular}
    
    \begin{tabular}{|c|c|c|}
    \hline
    	$\pidc\delsigsq$ & $\pidc\sigd^2$ & $(\delpic\cdot\delsigma)\dot{\sigma}$ \\
    	\hline
    	{\color{olive}$-2d_7/ f_\pi^2$} & {\color{magenta}$-2d_5/ f_\pi^2$} & {\color{orange}$-2d_4/ f_\pi^2$}\\
    	\hline
    \end{tabular}
    
    \begin{tabular}{|c|c|}
    \hline
    	$\sigd^3$ & $\delsigsq\sigd$ \\
    	\hline
    	{\color{violet}$-d_{9}/f_\pi^2$} & {\color{teal}$-d_{10}/f_\pi^2$}\\
    	\hline
    \end{tabular}
    \caption{Cubic interactions among the canonically normalized inflaton perturbation $\pi_{\rm c}$ and the spectator field perturbation $\sigma$, following from \cref{eq:IM_first}.
    Given the symmetry structure, some of the coefficients controlling the cubic interactions also control the quartic interactions among these perturbations.
    To illustrate this, we have used the same color to denote specific coefficients appearing both here and in \cref{tab:quartic_diff_basis}.}
    \label{tab:cubic_diff_basis}
    \end{center}
\end{table}
\begin{table}[!htb]
    \begin{center}
    \begin{tabular}{|c|c|c|}
    \hline
     $(\delpic)^4$ & $\pidc^2(\delpic)^2$ & $\pidc^4$ \\
    \hline
    {\color{red}$\frac{1}{2}M_2^4/f_\pi^8$} & {\color{blue}$2M_3^4/f_\pi^8$} & $\frac{2}{3}M_4^4/f_\pi^8$\\
    \hline
    \end{tabular}
    
    \begin{tabular}{|c|c|c|c|}
    \hline
     $\delpisqc(\delpic\cdot\delsigma)$ & $\pidc^2(\delpic\cdot\delsigma)$ & $\pidc\sigd\delpisqc$ & $\pidc^3\sigd$ \\
    \hline
    {\color{brown}$d_1 /f_\pi^4$} & {\color{cyan}$4d_2/ f_\pi^4$} & {\color{cyan}$4d_2/ f_\pi^4$} & $8d_3/ f_\pi^4$\\
    \hline
    \end{tabular}
    
    \begin{tabular}{|c|c|c|c|c|c|}
    \hline
     $(\delpic\cdot\delsigma)^2$ & $\pidc\sigd(\delpic\cdot\delsigma)$ & $\delpisqc\sigd^2$ & $\pidc^2\sigd^2$ & $\delpisqc\delsigsq$  & $\pidc^2\delsigsq$  \\
    \hline
    {\color{orange}$d_4/f_\pi^4$} & {\color{magenta}$4d_5/f_\pi^4$} & {\color{magenta}$d_5/f_\pi^4$} & $4d_6/f_\pi^4$ & {\color{olive}$d_7/f_\pi^4$} &  $4d_8/f_\pi^4$\\
    \hline
    \end{tabular}
    \begin{tabular}{|c|c|c|c|c|}
    \hline
     $\sigd^2(\delpic\cdot\delsigma)$ & $(\delpic\cdot\delsigma)\delsigsq$ & $\pidc\sigd^3$ & $\pidc\sigd\delsigsq$ \\
    \hline
    {\color{violet}$3d_{9}/f_\pi^4$} & {\color{teal}$d_{10}/f_\pi^4$} & $2d_{11}/f_\pi^4$ & $2d_{12}/f_\pi^4$\\
    \hline
    \end{tabular}
    
    \begin{tabular}{|c|c|c|}
    \hline
     $\sigd^4$ & $\sigd^2\delsigsq$ & $(\partial\sigma)^4$ \\
    \hline
    $d_{13}/f_\pi^4$ & $d_{14}/f_\pi^4$ & $d_{15}/f_\pi^4$\\
    \hline
    \end{tabular}
    \caption{Quartic interactions among the canonically normalized inflaton perturbation $\pi_{\rm c}$ and the spectator field perturbation $\sigma$, following from \cref{eq:IM_first}.
    As in \cref{tab:cubic_diff_basis}, we have used the same color to denote specific coefficients that appear both here and in \cref{tab:cubic_diff_basis}.}
    \label{tab:quartic_diff_basis}
    \end{center}
\end{table}

\subsubsection{Lorentz-Invariant EFT of Multifield Inflation in the Slow-Roll Regime}\label{sec:mult_field_sr}

The EFT described by ${\cal L}_{\rm IM}$ is quite general.
However, this generality comes at a cost in the form of a large number of coefficients to consider.
Given this profusion, it is convenient to also consider a simplified EFT that accounts for the homogeneous component of the inflaton $\phi$ that drives the inflationary expansion, along with the spectator field $\sigma$.
This will allow us to reduce the large parameter space while still retaining enough structure such that nontrivial consequences from positivity bounds will appear.
We begin with an EFT that contains the following dimension-eight operators at the leading order in derivatives in increasing powers of $\sigma$,
\begin{equation}
\label{eq:UVEFT}
  \begin{split}
    \mathcal{L}_\mathrm{UV}
      &= -\frac{1}{2}(\partial\phi)^2 -\frac{1}{2}\delsigsq
         + \frac{\kappa_1}{\Lambda^4} (\partial_\mu\phi)^2 (\partial_\nu\phi)^2
         + \frac{\kappa_2}{\Lambda^4} (\partial_\mu\phi)^2
                                      (\partial^\nu\phi\,\partial_\nu\sigma) \\
      &\qquad + \frac{\kappa_3}{\Lambda^4} (\partial_\mu\phi)^2 (\partial_\nu\sigma)^2
                + \frac{\kappa_4}{\Lambda^4} (\partial_\mu\phi\, \partial^\mu\sigma)^2 
                + \frac{\kappa_5}{\Lambda^4} (\partial^\mu\phi\, \partial_\mu\sigma)
                                             (\partial_\nu\sigma)^2
                + \frac{\kappa_6}{\Lambda^4} (\partial_\mu\sigma)^2 (\partial_\nu\sigma)^2,
  \end{split}
\end{equation}
where the $\kappa_i$ terms are various dimensionless couplings. 
We label this Lagrangian as ${\cal L}_{\rm UV}$, as this is a UV extension of the previous Lorentz-violating EFT~\eqref{eq:L_IM}, restoring Lorentz invariance above $|\dot{\phi}_0|^{1/2}$.

Manifestly, ${\cal L}_{\rm UV}$ has a significantly reduced number of interactions than ${\cal L}_{\rm IM}$, but we can relate the two descriptions.
To do so, we decompose $\phi$ into a spatially homogeneous and fluctuating component, $\phi(t,\vec{x}) = \phi_0(t) + \xi(t,\vec{x})$.
The fluctuations encoded in $\xi(t,\vec{x})$ are related to the $\pi(t,\vec{x})$ considered above by a multiplicative factor, $\xi \equiv \pi_{\rm c} = f_\pi^2\pi$ with $f_\pi^2 = \dot{\phi}_0$ for $\cs\simeq 1$.
The interaction terms at different orders are given by
\begin{equation}
\label{eq:brokenUVEFT}
  \begin{split}
    \mathcal{L}_\mathrm{UV} &\supset  
      \frac{\phid}{\Lambda^4}
        \bqty{-(4\kappa_1\dot{\xi} + \kappa_2 \sigd)(\partial\xi)^2
      	      - 2(\kappa_2\dot{\xi} + \kappa_4 \sigd)(\partial^\mu\xi\partial_\mu\sigma)
      	      - (2\kappa_3\dot{\xi} + \kappa_5\sigd)(\partial\sigma)^2} \\
&\qquad + \frac{1}{\Lambda^4}\left[\kappa_1(\partial\xi)^4 + \kappa_2(\partial\xi)^2(\partial^\mu\xi\partial_\mu\sigma) + \kappa_3 (\partial\xi)^2(\partial\sigma)^2
 + \kappa_4 (\partial^\mu\xi\partial_\mu\sigma)^2 \right. \\ & \qquad \qquad \qquad \left. + \kappa_5 (\partial^\mu\xi \partial_\mu\sigma)(\partial\sigma)^2 + \kappa_6(\partial\sigma)^4\right].
  \end{split}
\end{equation}

As written, the EFT in \cref{eq:brokenUVEFT} can be compared to the Goldstone EFT by noting that $\delta g^{00}$ is related to $(\partial\phi)^2$ in the slow-roll approximation,
\begin{equation}
  (\partial\phi)^2 = \phid^2 \bqty{- 1 - 2\pid + \delpisq},
\end{equation}
and $g^{0\mu}\partial_\mu\sigma$ is related to $\partial^\mu\sigma\partial_\mu\phi$ by
\begin{equation}
  \partial^\mu\sigma\partial_\mu\phi
    = \phid \pqty{-\sigd - \pid\sigd + \frac{1}{a^2}\delipi \delisigma}.    
\end{equation}
With this connection between the two notations, we see that among all the $M_n$ and $d_i$ coefficients, only six correspond to dimension-eight operators in the Lorentz-invariant UV theory.
After matching all such cubic and quartic interactions, the theories are related by
\begin{equation}
\label{eq:UVIREFTrelations}
  M_2^4 = 2\kappa_1 \frac{f_\pi^8}{\Lambda^4} \qc
  d_1 = \kappa_2 \frac{f_\pi^4}{\Lambda^4} \qc
  d_4 = \kappa_4 \frac{f_\pi^4}{\Lambda^4} \qc
  d_7=\kappa_3 \frac{f_\pi^4}{\Lambda^4} \qc
  d_{10} = \kappa_5 \frac{f_\pi^4}{\Lambda^4} \qc
  d_{15} = \kappa_6 \frac{f_\pi^4}{\Lambda^4}.
\end{equation}
In other words, the remaining Wilson coefficients appearing in $\mathcal{L}_\mathrm{IM}$ must arise from operators in this particular Lorentz-invariant UV EFT with dimension higher than eight.
For example, the operator $d_{13}\sigd^4$ would arise independently from $(\partial\sigma)^4$, and in fact originates from a dimension-sixteen term, $(\partial\phi\cdot\partial\sigma)^4$. 
In this sense, this simple UV EFT selects a leading six-dimensional coefficient space where we can place useful positivity bounds from analyticity and the generalized optical theorem.

Following the same logic as in \cref{sec:iso_multifield}, we have used the same EFT cutoff scale $\Lambda$ for both the $\phi$ and the $\sigma$ sector. Since the homogeneous inflaton dynamics is a part of the EFT description, we have $\Lambda > |\dot{\phi}_0|^{1/2}$. This hierarchy of scales is summarized in \cref{fig:scales}~(d).

\section{Full Bounds}
\label{sec:bounds}

Drawing on the four classes of EFTs described in \cref{sec:infeft}, we can now compute various $2 \to 2$ amplitudes involving $\pi_{\rm c}$ and $\sigma$.
With these amplitudes in hand, we can then apply the positivity bounds derived in \cref{sec:gen_opt_thm} to obtain constraints on the corresponding EFTs.
The bounds that result can then be applied to the parameter space considered in searches for NG, as we will demonstrate in \cref{sec:obs}.

\subsection{Warm-up: Positivity Bounds on \texorpdfstring{$\mathcal{L}_\mathrm{AS}$}{the AS Term}}

We start by considering the EFT in \cref{eq:LAS_final} with $\cs\simeq 1$, an example that was considered in \Refc{Grall:2021xxm}, and we will reproduce their results below.
The theory contains a single field, and the physical process relevant for deriving positivity bounds is $\pi_{\rm c}\pi_{\rm c}\to \pi_{\rm c}\pi_{\rm c}$ scattering.
To simplify the notation, as in \Refc{Baumann:2015nta} we define:
\bea
\alpha_1 &= -8c_2^2-\frac{4}{3}c_3 \qquad\qquad &\beta_1&=8c_2^3+8c_2c_3+\frac{2}{3}c_4\\
\alpha_2 &= 2c_2 & \beta_2 &= -2c_3-4c_2^2 & \\ && \beta_3&=\frac{c_2}{2}.
\eea
To express the amplitude in terms of $\{\omega_s,\omega_t,\omega_u,s,t\}$, we relate the energies,
\bea
\omega_1 &= \frac{1}{2} (\omega_s + \omega_t + \omega_u), & \qquad  \omega_2 &= \frac{1}{2} (\omega_s - \omega_t - \omega_u), \\
\omega_3 &= \frac{1}{2} (-\omega_s + \omega_t - \omega_u),& \qquad  \omega_4 &= \frac{1}{2} (-\omega_s - \omega_t + \omega_u),
\eea
and use the Breit parameterization~\eqref{eq:BreitParam},
\be
\omega_s = \frac{1}{8M}(s - u + 8 M^2 \gamma^2),\hspace{0.5cm}
\omega_u = \frac{1}{8M}(-s + u + 8 M^2 \gamma^2).
\ee
We will further make use of energy conservation $\sum_i\omega_i=0$, along with $s+t+u=0$ as appropriate for the scattering of massless particles.

Using these relations, we can compute the forward limit of the $\pi_{\rm c}\pi_{\rm c}\to \pi_{\rm c}\pi_{\rm c}$ scattering amplitude by taking $\omega_t\to0$ and then $t\to 0$.
The result is given by
\be
\mathcal{M}_{\pi_{\rm c}\pi_{\rm c}\to\pi_{\rm c}\pi_{\rm c}}(s) = \frac{s^2}{4 f_\pi^4}\left[16 \beta_3 + 8(\beta_2 - 2 \alpha_2^2) \gamma^2 + 6(\beta_1 - 4 \alpha_1 \alpha_2) \gamma^4 - 9 \alpha_1^2 \gamma^6\right]\!.
\ee
Demanding $\mathcal{M}_{\pi_{\rm c}\pi_{\rm c}\to\pi_{\rm c}\pi_{\rm c}}''(s)>0$ then gives the positivity bound,
\be
16 \beta_3 + 8(\beta_2 - 2 \alpha_2^2) \gamma^2 + 6(\beta_1 - 4 \alpha_1 \alpha_2) \gamma^4 - 9 \alpha_1^2 \gamma^6 > 0.	
\ee
As noted in \Refc{Grall:2021xxm}, for $\gamma=1$ this reproduces the bound from \Refc{Baumann:2015nta}.
However, this is unsurprising.
As explained in \cref{sec:twofield}, $\gamma=1$ corresponds to the CM frame, within which the calculations of \Refc{Baumann:2015nta} were performed.

We can rewrite the bound in terms of the coefficients $c_i$ as
\be
2 c_2 - 4 (6 c_2^2 + c_3) \gamma^2 + 
(108 c_2^3 + 28 c_2 c_3 + c_4) \gamma^4 - 
4 (6 c_2^2 + c_3)^2 \gamma^6 > 0.
\ee
In particular, demanding $\cs\simeq 1$ forces $|c_2|\ll |c_3|,|c_4|$, and correspondingly the positivity bound simplifies to
\be
c_4 \gamma^2 > 4 c_3 (1 + c_3 \gamma^4).
\ee

\subsection{Positivity Bounds on \texorpdfstring{$\mathcal{L}_\mathrm{AM}$}{the AS Term}}

Next, we take the theory where adiabatic fluctuations arise from a field $\chi$ that is distinct from the inflaton, as described by \cref{eq:LAM}.
Taking the amplitude for $\chi \chi \to \chi \chi$ scattering in the forward limit yields
\be
\mathcal{M}_{\chi\chi\to\chi\chi} = \frac{s^2}{2\Lambda_\chi^4} \left[8 e_3 - 4 (e_2 + 2 e_3) \gamma^2 + 3 (e_1 + e_2 + e_3) \gamma^4\right].	
\ee
As before, we demand $\mathcal{M}''_{\chi\chi\to\chi\chi}(s)>0$ to arrive at the positivity bound,
\be\label{eq:LAM_bound}
8 e_3 - 4 (e_2 + 2 e_3) \gamma^2 + 3 (e_1 + e_2 + e_3) \gamma^4 > 0.	
\ee
A consequence of the assumed $\ZZ_2$ symmetry acting as $\chi\to-\chi$ is that this bound involves only three coefficients.
Furthermore, as we will see in \cref{sec:obs}, two of these coefficients give overlapping trispectrum shapes.
Therefore, this positivity constraint allows for a powerful restriction on the two-dimensional trispectrum parameter space.

\subsection{Positivity Bounds on \texorpdfstring{$\mathcal{L}_\mathrm{IM}$}{the IM term}}

We next consider the two-field theory that can generate isocurvature perturbations, as described in \cref{eq:IM_first,eq:L_IM} and summarized in \cref{tab:cubic_diff_basis,tab:quartic_diff_basis}.
Invariably, the associated positivity bounds will involve more parameters.
We focus on the bounds that can be derived from the conventional optical theorem, so that we start with the computation of all two-to-two scatterings with an elastic forward limit: $\pi_{\rm c}\pi_{\rm c} \to \pi_{\rm c}\pi_{\rm c}$, $\sigma\sigma \to \sigma\sigma$, and $\pi_{\rm c}\sigma \to \pi_{\rm c}\sigma$.
Along the way, we will also briefly mention which of the Wilson coefficients can be potentially constrained using the current Planck data.
At the same time, we will describe the observables that, if constrained, would allow us to impose all the derived positivity bounds.

\paragraph{$\boldsymbol{\pi}_\mathbf{c}\boldsymbol{\pi}_\mathbf{c} \boldsymbol{\to} \boldsymbol{\pi}_\mathbf{c}\boldsymbol{\pi}_\mathbf{c}$ scattering.}
The relevant vertices that generate this amplitude are given by two $\pi_{\rm c}^3$ interactions (suppressing derivatives) with a $\pi_{\rm c}$ exchange, two $\pi_{\rm c}^2\sigma$ interactions with a $\sigma$ exchange, and the contact $\pi_{\rm c}^4$ interactions.
The positivity bound from the scattering amplitude in the forward limit is given by
\be
4 c_2
- 8 (c_3 + 4 c_2^2 + d_1^2) \gamma^2
+ 2 (c_4 + 16 c_2 c_3 + 8 d_1 d_2) \gamma^4
- 8 (c_3^2 + d_2^2) \gamma^6
> 0.
\label{eq:pi2to2}
\ee
We note that $c_2, c_3, c_4$ control the adiabatic three- and four-point functions, and $d_1, d_2$ control three-point functions involving one isocurvature and two adiabatic fluctuations.
Since both classes of NG have been searched for using Planck data, albeit with the local shape, the above bound can be imposed on the observational parameter space.

\paragraph{$\boldsymbol{\sigma}\boldsymbol{\sigma} \boldsymbol{\to} \boldsymbol{\sigma}\boldsymbol{\sigma}$ scattering.}
Similarly to above, in this case the relevant vertices are given by two $\sigma^3$ interactions with a $\sigma$ exchange, two $\sigma^2\pi_{\rm c}$ interactions with a $\pi_{\rm c}$ exchange, and contact $\sigma^4$ interactions.
Correspondingly, the positivity bound is given by
\bea
16 d_{15} 
- 4 (4 d_{10}^2 + 2 d_{14} + [d_4 + 2 d_7]^2) \gamma^2& \\
+ 2 (3 d_{13} + 4 d_4 d_5 + 8 d_5 d_7 + 12 d_9 d_{10}) \gamma^4
- (4 d_5^2 + 9 d_9^2) \gamma^6
&> 0.
\label{eq:sig2to2}
\eea
We note that if $\pi_{\rm c}$ is absent from the spectrum and we were to impose a $\ZZ_2$ symmetry $\sigma\to-\sigma$, then the above constraint reduces to
\be
8 d_{15}-4 d_{14}\gamma^2+3 d_{13}\gamma^4 > 0.
\ee
This exactly matches the bound in \cref{eq:LAM_bound} upon identifying $d_{13}\to e_1+e_2+e_3$, $d_{14}\to e_2+2e_3$, and $d_{15}\to e_3$.
We can further derive the positive bound in the absence of $\pi_{\rm c}$, but without imposing the $\ZZ_2$ on $\sigma$, in which case \cref{eq:sig2to2} reduces to
\be
16 d_{15} 
- 8 (2 d_{10}^2 + d_{14}) \gamma^2
+ 6 (d_{13} + 4 d_9 d_{10}) \gamma^4
- 9 d_9^2 \gamma^6
> 0.
\ee
The coefficient $d_{15}$ only controls the isocurvature four-point function, which has not yet been observationally constrained.
Therefore we cannot apply this positivity bound directly.

\paragraph{$\boldsymbol{\pi}_\mathbf{c} \boldsymbol{\sigma} \boldsymbol{\to} \boldsymbol{\pi}_\mathbf{c}\boldsymbol{\sigma}$ scattering.}
For the final amplitude, in addition to contact terms, the scattering can now be mediated by either a $\pi_{\rm c}$ or $\sigma$ exchange.
Accounting for all possibilities, the positivity bound is
\be
d_4
- 2(2d_1^2 + d_4^2 + d_5 + 2 d_4 d_7) \gamma^2
+ (8d_1 d_2 + 3 d_4 d_5 + d_6 + 2 d_5 d_7) \gamma^4
- (4d_2^2 + d_5^2) \gamma^6
>0.
\label{eq:pisigtopisig}
\ee
In a similar vein as above, we note that the coefficient $d_6$ controls only a four-point function involving two adiabatic and two isocurvature modes.
This interaction has also not been searched for in the data.
Therefore, once again we cannot apply the positivity bound directly.

The relations in \cref{eq:pi2to2,eq:sig2to2,eq:pisigtopisig} involve fourteen of the eighteen parameters that describe $\mathcal{L}_\mathrm{IM}$.
In particular, $d_3$, $d_8$, $d_{11}$, and $d_{12}$ appear nowhere.
Of these, $d_8$ does in fact mediate $\pi_{\rm c}\sigma \to \pi_{\rm c}\sigma$, but the corresponding amplitude is proportional to Mandelstam $t$ and therefore vanishes in the forward limit, so bounding it would require beyond-forward methods, e.g., Refs.~\cite{deRham:2017imi, Bellazzini:2017fep}.
The remaining three missing $d_i$ are contact interactions with an odd number of $\pi_{\rm c}$ and $\sigma$, and therefore do not contribute to elastic scattering.
In principle, we could access these coefficients by drawing on the power of the generalized optical theorem.
Instead, in the next subsection, we will highlight the power of the generalized optical theorem by considering the simplified version $\mathcal{L}_\mathrm{IM}$ of introduced in \cref{sec:mult_field_sr}.

\subsection{Positivity Bounds on \texorpdfstring{$\mathcal{L}_\mathrm{UV}$}{the UV term}}

Finally, we consider the reduction of $\mathcal{L}_\mathrm{IM}$ introduced in \cref{sec:mult_field_sr}.
While $\mathcal{L}_\mathrm{UV}$ may be less general, its structure affords us the opportunity to study several additional aspects of the positivity bounds.
First, it allows us to consider bounds from the generalized optical theorem and to demonstrate explicitly that they are stronger than those accessible purely from elastic forward amplitudes.
Second, as we know a UV extension of \cref{eq:brokenUVEFT}---indeed, we derived it directly from one---we can study explicitly the fate of our bounds in the Lorentz-invariant UV.

\paragraph{Positivity in the UV.}
To begin, we consider scattering with $E\gg |\dot{\phi}_0|^{1/2}$.
Accordingly, we can work with the full inflaton field $\phi$, and the appropriate description is the Lagrangian in \cref{eq:UVEFT}.
Interactions in the theory are mediated solely  by contact terms, and the positivity bounds simplify accordingly.

If we study $\phi\phi\to\phi\phi$, $\sigma\sigma\to\sigma\sigma$, and $\phi\sigma\to\phi\sigma$, then the three bounds are
\be
\kappa_1 > 0,\hspace{0.5cm}
\kappa_6 > 0,\hspace{0.5cm}
\kappa_4 > 0,
\label{eq:kappa_164}
\ee
respectively.
Note that while the amplitude for purely $\phi$ and $\sigma$ scattering only depend on a single Wilson coefficient, for $\phi\sigma$ scattering the amplitude is
\be
\mathcal{M}_{\phi\sigma\to\phi\sigma} = \frac{1}{\Lambda^4}\left[\kappa_3 t^2 + \frac{1}{2} \kappa_4 (s^2 + u^2) \right]\!.
\ee
We see the appearance of $\kappa_3$, but as it only enters proportional to $t^2$, it cannot be constrained from a forward amplitude.

\paragraph{Positivity in the IR.}
We next consider the same theory at lower energies.
In particular, we imagine scattering with external states having energies that satisfy $H \ll E \ll |\dot{\phi}_0|^{1/2}$.
To this end, we can write $\phi(t,\vec{x})=\phi_0(t)+\xi(t,\vec{x})$, which gives a set of cubic and quartic interactions that were summarized in \cref{eq:brokenUVEFT}.
We can then use the bounds on $\mathcal{L}_\textrm{IM}$ derived in \cref{eq:pi2to2,eq:sig2to2,eq:pisigtopisig}, which when keeping only the terms that do not vanish in $\mathcal{L}_\textrm{UV}$ become
\begin{equation}
  c_2 > 2(4 c_2^2 + d_1^2) \gamma^2 \qc
  d_{15} > \pqty{d_{10}^2 + \frac{1}{4}(d_4 + 2 d_7)^2 } \gamma^2 \qc
  d_4 > 2(2d_1^2 + d_4^2 + 2 d_4 d_7) \gamma^2.
\end{equation}
After applying the translation given in \cref{eq:UVIREFTrelations}, these bounds become
\begin{equation}
\label{eq:kappa_164_IR}
  \begin{split}
    \kappa_1 &> \frac{\gamma^2 f_\pi^4}{\Lambda^4} (16 \kappa_1^2 + \kappa_2^2) \\
    \kappa_6 &> \frac{\gamma^2 f_\pi^4}{\Lambda^4}
                  \pqty{\kappa_5^2 + \frac{1}{4} (\kappa_4 + 2 \kappa_3)^2 } \\
    \kappa_4 &> \frac{\gamma^2 f_\pi^4}{\Lambda^4}
                  (4\kappa_2^2 + 2 \kappa_4^2 + 4 \kappa_3\kappa_4).
  \end{split}
\end{equation}
The expression on the right-hand side of the first two relations in \cref{eq:kappa_164_IR} are positive definite, implying that the bounds on $\kappa_1$ and $\kappa_6$ are strictly stronger than those in \cref{eq:kappa_164}, which were derived in the UV.
Initially, this result may seem surprising.
It suggests the consistency conditions encoded in positivity become stronger at low energies, and therefore a consistent theory in the UV could flow to an inconsistent one in the IR.

As a first point, we note that for $\gamma \sim 1$, the corrections to the UV bounds are parametrically small.
The validity of the UV EFT requires $\Lambda \gg |\dot{\phi}_0|^{1/2} \simeq f_\pi\simeq 60H$,\footnote{The equation of motion for the homogeneous inflaton field implies $\dot{\phi}_0^2 = 2\Mpl^2\dot{H}\simeq f_\pi^4$, for $c_s\simeq
 1$.} so that effects at $\mathcal{O}(f_\pi^4/\Lambda^4)$ are small.
This is to be expected since in \cref{eq:UVEFT} we have truncated the EFT expansion at mass dimension eight.
We note that if we parametrically increase $\gamma$ beyond an ${\cal O}(1)$ value, this can be interpreted as lowering the cutoff of the EFT to $\Lambda\to \Lambda/\gamma^n$, where the precise power $n$ is dependent on the derivative structure of the terms that contribute to the amplitude.
This can be understood from the origin of $\gamma$ in the Breit parameterization in \cref{eq:BreitParam} as follows. In the forward limit $4 \omega_1 \omega_2 = \gamma^2 s$.
Therefore, even for fixed $s$, increasing $\gamma$ implies that the states have energies larger than $\sqrt{s}$.
Correspondingly, whenever these energies become of order the EFT cutoff, the EFT can break down even if $s\ll \Lambda^2$.
Therefore, EFT consistency suggests that we do not expect the IR corrections in \cref{eq:kappa_164_IR} to be parametrically large.

Given that the correction is expected to be parametrically small, we now consider whether it should be satisfied in the UV.
Consider the first bound in \cref{eq:kappa_164_IR} derived from $\phi\phi\to\phi\phi$ scattering.
As a first possibility, imagine that we are studying a theory where there is an additional $\ZZ_2$ symmetry satisfied by $\sigma$, so that we can forbid the interaction given by $\kappa_2$.
In such an event, the bound becomes
\begin{equation}
  \kappa_1 \pqty{ 1 - \frac{\gamma^2 f_\pi^4}{\Lambda^4} 16 \kappa_1 } > 0,
\end{equation}
or equivalently,
\begin{equation}
\label{eq:kappa-2sided}
  0 < \kappa_1 < \frac{\Lambda^4}{16 \gamma^2 f_\pi^4}.
\end{equation}
The additional IR condition is translated into an upper bound on $\kappa_1$.
However, perturbative unitarity for the EFT in $\mathcal{L}_\mathrm{UV}$ already imposes the condition that $\kappa_1$ cannot be parametrically larger than unity, which is what would be required to violate the condition \cref{eq:kappa-2sided}.
Accordingly, the condition is straightforwardly satisfied in the UV.
If we restore $\kappa_2$ to the UV, then one could imagine a scenario where we tune $\kappa_1 = 0$ but $\kappa_2 \neq 0$, and a violation of the first constraint in \cref{eq:kappa_164_IR} appears possible.
The violation, however, would occur at $\mathcal{O}(\kappa^2)$, and at this same order we can generate $\kappa_1$ at one loop via a bubble diagram involving a $\pi$ and $\sigma$ in the loop.
The diagram is UV divergent, but if we cut the loop momenta off at $\Lambda$ then the divergence will exactly compensate the suppression we get from having an additional dimension-eight interaction.
Parametrically, we then expect to generate $\kappa_1 \sim \kappa_2^2/16\pi^2$, so that even in this tuned scenario, so long as $1/16\pi^2 > \gamma^2 f_{\pi}^4/\Lambda^4$, the UV will again satisfy the IR bounds.

For the reasons above, it appears that the additional constraints that appear in the IR are in fact straightforwardly satisfied in the UV.
We emphasize that this does not imply the restrictions are without content.
The bounds in \cref{eq:kappa_164_IR} were derived without any knowledge of the UV.
Further, we were able access information in the IR using tree-level calculations that are only satisfied in the UV once loops are included.
That this occurred can be traced back to the symmetry breaking at the heart of the nonrelativistic theories: once we have a background value of $\phi$ to expand around, we can convert quartic contact interactions to three-point operators, and therefore allow new interactions to contribute to the forward amplitude.
The ability to vary the order of the interactions and therefore the constraints appears to be one of the key powers of this approach.

\paragraph{Extension to include the generalized optical theorem.}
As an example of the power of the generalized optical theorem, here we demonstrate how our UV positivity bounds $\kappa_1, \kappa_4, \kappa_6 > 0$ can be extended using the formalism derived in \cref{sec:gen_opt_thm}.
We first compute the associated $M_{ijkl}$. Based on the EFT in \cref{eq:UVEFT}, they are given by
\begin{equation}
  \begin{split}
    \amp_{1111} = \frac{2\kappa_1}{\Lambda^4} (s^2 + t^2 + u^2)
      &\to M_{1111} = \frac{8\kappa_1}{\Lambda^4}\\
	\amp_{1112} = \frac{\kappa_2}{2\Lambda^4} (s^2 + t^2 + u^2)
	  &\to M_{1112} = \frac{2\kappa_2}{\Lambda^4}\\
	\amp_{1122} = \frac{\kappa_3}{\Lambda^4} s^2 + \frac{\kappa_4}{2\Lambda^4} (t^2 + u^2)
	  &\to M_{1122} = \frac{2\kappa_3+\kappa_4}{\Lambda^4}\\
	\amp_{1212} = \frac{\kappa_3}{\Lambda^4} t^2 + \frac{\kappa_4}{2\Lambda^4} (s^2 + u^2)
	  &\to M_{1212} = \frac{2\kappa_4}{\Lambda^4}\\
	\amp_{1222} = \frac{\kappa_5}{2\Lambda^4} (s^2 + t^2 + u^2)
	  &\to M_{1222} = \frac{2\kappa_5}{\Lambda^4}\\
	\amp_{2222} = \frac{2\kappa_6}{\Lambda^4} (s^2 + t^2 + u^2)
	  &\to M_{2222} = \frac{8\kappa_6}{\Lambda^4}.
  \end{split}
\end{equation}
The positivity bounds in \cref{eq:gtrzero} coincide with those in \cref{eq:kappa_164}, as they should.
Therefore, we focus on the remaining bounds consisting of \cref{eq:M1212mu,eq:cos,eq:triangle}.
For simplicity we define, $\tilde{\kappa}_i = \kappa_i/\Lambda^4$.
Then \cref{eq:M1212mu} gives
\begin{equation}
	\tilde{\kappa}_4 > \frac{\mu}{4}. 
\end{equation}
Next, \cref{eq:cos} yields
\begin{equation}
  \begin{split}
    (4 \tilde{\kappa}_3  + \mu)^2 &< 64\tilde{\kappa_1}\tilde{\kappa_6} \\
    \tilde{\kappa}^2_2 &< \tilde{\kappa}_1\left(4\tilde{\kappa}_4 - \mu\right) \\
	\tilde{\kappa}_5^2 &< \tilde{\kappa}_6\left(4\tilde{\kappa}_4 - \mu\right).
  \end{split}
\end{equation}
Finally the triangle inequality \cref{eq:triangle} requires
\begin{equation}
  \begin{split}
    \sqrt{1-\frac{4\tilde{\kappa}_3+\mu}{8\sqrt{\tilde{\kappa}_1\tilde{\kappa}_6}}}
      &< \sqrt{1-\frac{2\tilde{\kappa}_2}{\sqrt{8\tilde{\kappa}_1\left(2\tilde{\kappa}_4-\frac{1}{2}\mu\right)}}}
         +\sqrt{1-\frac{2\tilde{\kappa}_5}{\sqrt{8\tilde{\kappa}_6\left(2\tilde{\kappa}_4-\frac{1}{2}\mu\right)}}} \\
    \sqrt{1-\frac{2\tilde{\kappa}_2}{\sqrt{8\tilde{\kappa}_1\left(2\tilde{\kappa}_4-\frac{1}{2}\mu\right)}}}
      &<\sqrt{1-\frac{2\tilde{\kappa}_5}{\sqrt{8\tilde{\kappa}_6\left(2\tilde{\kappa}_4-\frac{1}{2}\mu\right)}}}
         +\sqrt{1-\frac{4\tilde{\kappa}_3+\mu}{8\sqrt{\tilde{\kappa}_1 \tilde{\kappa}_6}}}\\
    \sqrt{1-\frac{2\tilde{\kappa}_5}{\sqrt{8\tilde{\kappa}_6\left(2\tilde{\kappa}_4-\frac{1}{2}\mu\right)}}}
      &< \sqrt{1-\frac{4\tilde{\kappa}_3+\mu}{8\sqrt{\tilde{\kappa}_1 \tilde{\kappa}_6}}}
           +\sqrt{1-\frac{2\tilde{\kappa}_2}{\sqrt{8\tilde{\kappa}_1\left(2\tilde{\kappa}_4-\frac{1}{2}\mu\right)}}}.
  \end{split}
\end{equation}
An equivalent condition to these inequalities can be written as
\bea
\label{eq:UVbound_gen_opt}
 &1+\frac{(4\kappa_3+\mu)\left(\kappa_2\sqrt{2\kappa_6}+\kappa_5\sqrt{2\kappa_1}\right)\sqrt{\left(2\kappa_4-\frac{1}{2}\mu\right)}}{16\kappa_1 \kappa_6 \left(2\kappa_4-\frac{1}{2}\mu\right)} 
+\frac{\kappa_2 \kappa_5 \sqrt{\kappa_1 \kappa_6}}{2\kappa_1 \kappa_6 \left(2\kappa_4-\frac{1}{2}\mu\right)}\\
& > \frac{1}{4}\left[1+\frac{4\kappa_3+\mu}{8\sqrt{\kappa_1 \kappa_6}}+\frac{\kappa_2}{\sqrt{2\kappa_1 \left(2\kappa_4 - \frac{1}{2}\mu\right)}}+\frac{\kappa_5}{\sqrt{2\kappa_6\left(2\kappa_4 -\frac{1}{2}\mu\right)}}\right]^{2}.
\eea
These additional constraints impose strictly stronger requirements on the theory, as we will demonstrate next.

\begin{figure}[!t]
\begin{center}
\includegraphics[width=14cm]{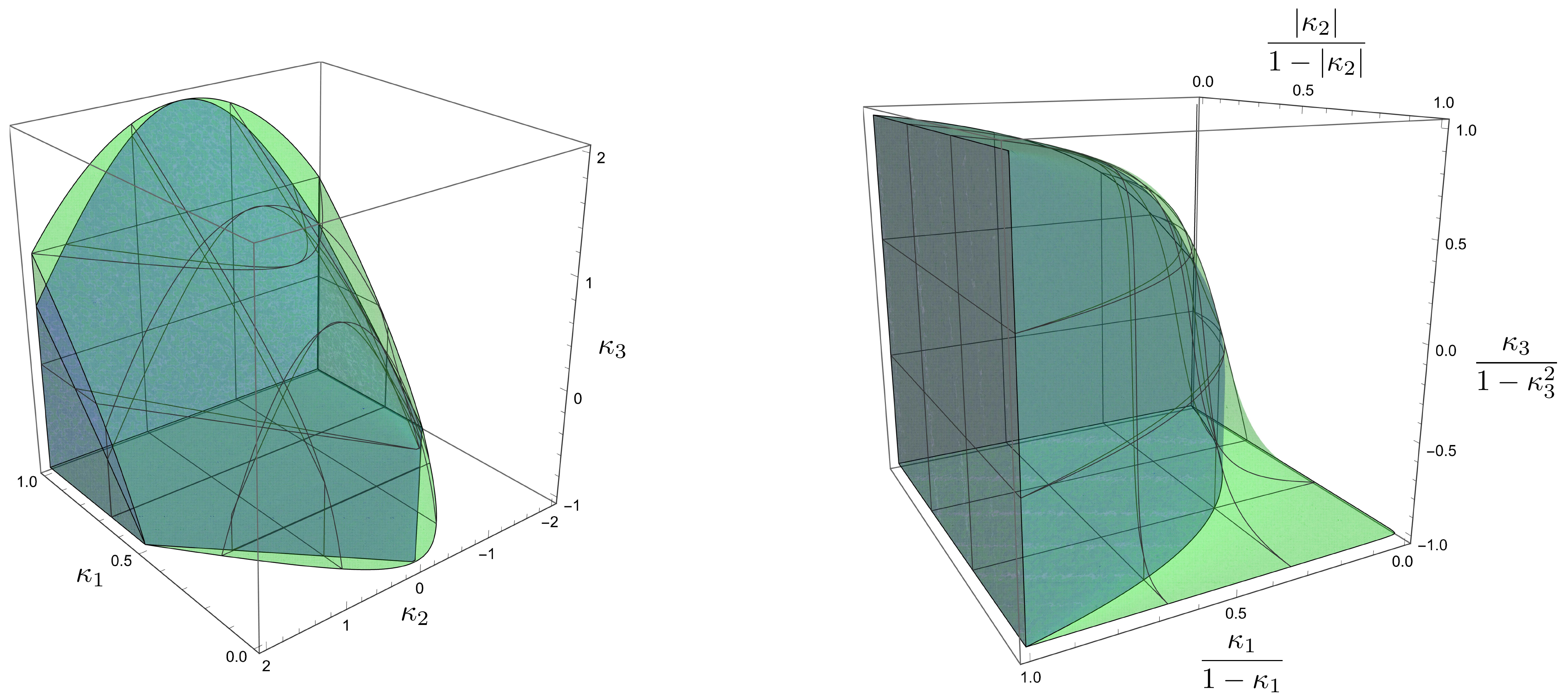}
\end{center}
\vspace{-0.5cm}
\caption{Region in the $(\kappa_1,\kappa_2,\kappa_3)$ parameter space, for a $D_4$-symmetric theory, permitted by the generalized optical theorem (blue), compared with the (larger) region permitted by considering merely elastic scattering of arbitrary superpositions (green).}
\label{fig:kappa3d}
\end{figure}

\paragraph{Theory with dihedral symmetry.}
The discussion above has focused on a theory with arbitrary interactions $\kappa_i$ among the spectator field $\sigma$ and inflaton $\phi$.
In order to visualize the impact of the generalized optical theorem, it is illustrative to reduce the parameter space to a more manageable number of Wilson coefficients, while still retaining the nontrivial structure we found in \cref{sec:gen_opt_thm}.
We can do so by considering a model in which the fundamental theory enjoys a set of discrete symmetries among $\phi$ and $\sigma$ generated by $(\phi,\sigma)\to (\sigma,-\phi)$ and $(\phi,\sigma)\to \left(\tfrac{\phi + \sigma}{\sqrt{2}},\tfrac{\phi-\sigma}{\sqrt{2}}\right)$.
In geometric language, these symmetries together generate the octic group, i.e., the order-eight dihedral group $D_4$.
While we can require that the full Lagrangian possess these symmetries, this does not prevent us from considering initial conditions that break them, namely, an inflaton vev in $\phi$ alone.
The dihedral symmetry requires three constraints among the six $\kappa_i$ of \cref{eq:UVEFT}, $\kappa_4 = 2\kappa_1 - \kappa_3$, $\kappa_5 = -\kappa_2$, and $\kappa_6 = \kappa_1$, leaving us with a three-dimensional parameter space.
The generalized optical theorem bound in \cref{eq:UVbound_gen_opt}, along with the weaker bound one would obtain from applying the standard optical theorem to the elastic scattering of arbitrary superpositions of $\phi$ and $\sigma$, is depicted in \cref{fig:kappa3d}.
Those we obtain from the generalized optical theorem are strictly stronger.

We could apply the same generalized optical theorem technology to the boost-breaking EFT as in \cref{eq:L_IM} without obstruction, and would obtain a similar qualitative strengthening beyond elastic positivity bounds, albeit in a much larger parameter space.

\section{Observational Constraints on non-Gaussianities}
\label{sec:obs}

Positivity bounds imply nontrivial restrictions on primordial NG since the same parameters $c_i,d_i,\kappa_i$ that we constrained in \cref{sec:bounds} also determine the strength of NG for various kinematic configurations. 
In Refs.~\cite{Baumann:2015nta,Grall:2021xxm}, such observational constraints were considered in the context of single-field inflation with adiabatic perturbations, including an exploration of the implications of positivity within $\mathcal{L}_\mathrm{AS}$.
However, as we have seen in \cref{sec:bounds}, once multiple fields are present, the bounds can change considerably, in addition to having relevance for probes of isocurvature perturbations.
We therefore focus our discussion in this section on scenarios where more than one field is present during inflation.

\subsection{Constraints on ${\cal L}_{\rm AM}$}

In this subsection we consider the EFT in \cref{eq:LAM} for which we effectively have a single source of fluctuations $\chi$ contributing to primordial perturbations.
Therefore, subtleties regarding causality and analyticity due to multiple sound speeds do not arise, and at the same time, our positivity bounds can be imposed on the parameter space allowed by current constraints on NG.

To compute the NG of the curvature perturbation $\zeta$ based on \cref{eq:LAM}, we use the linear relationship $\zeta = \sqrt{2 A_\zeta}\chi/H$ between $\zeta$ and $\chi$ fluctuations.
Here the normalization is fixed by the power spectrum $P_\zeta(k) = A_\zeta/k^3$.
The four-point functions sourced by $e_1,e_2, e_3$ can be computed as~\cite{Smith:2015uia}
\begin{align}
  \ev*{\zeta^4}' \rvert_{e_1}
    & = \frac{288 e_1 A_\zeta^2}{\Lambda_\chi^4}
          \frac{1}{k_t^5\prod_{i=1}^4 k_i} \nonumber \\
  \ev*{\zeta^4}' \rvert_{e_2}
    & = \frac{4 e_2 A_\zeta^2}{\Lambda_\chi^4}
          \bqty{\frac{k_t^2 + 3(k_3+k_4)k_t + 12 k_3 k_4}{k_1k_2k_3^3k_4^3 k_t^5}
          	      (\vk_3 \cdot \vk_4) + \text{5 perms.}}\\
  \ev*{\zeta^4}' \rvert_{e_3}
    & = \frac{8e_3 A_\zeta^2}{\Lambda_\chi^4}
          \bqty{\frac{2k_t^4 - 2k_t^2\sum_{i=1}^4 k_i^2
          		      + k_t \sum_{i=1}^4 k_i^3 + 12\prod_{i=1}^4 k_i}
          	         { k_t^5\prod_{i=1}^4 k_i^3}
          	      (\vk_1 \cdot \vk_2)(\vk_3 \cdot \vk_4) + \text{2 perms.}}. \nonumber
\end{align}
Here $k_t \equiv k_1+k_2+k_3+k_4$ with $k_i = |\vk_i|$, and the sums and products run over $i=1,2,3,4$. 
The quantity $\ev*{\zeta^4}'$ is defined by $\ev{ \zeta(\vk_1) \zeta(\vk_2) \zeta(\vk_3) \zeta(\vk_4)} \equiv \ev*{\zeta^4}' (2\pi)^3 \delta^3(\vk_1 + \vk_2 + \vk_3 + \vk_4)$.

The observational constraints from the Planck data can be described in terms of three parameters $g_\mathrm{NL}^{\dot{\sigma}^4}, g_\mathrm{NL}^{\dot{\sigma}^2(\partial\sigma)^2}$ and $g_\mathrm{NL}^{(\partial\sigma)^4}$.
These are related to $e_1, e_2, e_3$ as~\cite{Smith:2015uia},
\begin{equation}\label{eq:gNLs}
g_{\rm NL}^{\dot{\sigma}^4} A_\zeta =\frac{25}{768}\frac{e_1 H^4}{\Lambda_\chi^4},\qquad  g_{\rm NL}^{\dot{\sigma}^2(\partial\sigma)^2} A_\zeta  = -\frac{325}{6912}\frac{e_2H^4}{\Lambda_\chi^4},\qquad  g_{\rm NL}^{(\partial\sigma)^4} A_\zeta = \frac{2575}{20736}\frac{e_3H^4}{\Lambda_\chi^4}.
\end{equation}
However, the trispectrum shapes mediated by these three operators are correlated with each other. 
Therefore the parameter space can be described in terms of only two of these parameters, which can be chosen to be $g_{\rm NL}^{\dot{\sigma}^4}$ and $g_{\rm NL}^{(\partial\sigma)^4}$. The effect of $g_{\rm NL}^{\dot{\sigma}^2(\partial\sigma)^2}$ can then be absorbed to give rise to effective coefficients~\cite{Smith:2015uia, Planck:2019kim},
\begin{equation}
\label{eq:eff_coeff}
	g_\mathrm{NL}^{\dot{\sigma}^4}\rvert_\mathrm{eff}
	  = 0.597 g_\mathrm{NL}^{\dot{\sigma}^2(\partial\sigma)^2} \qc
	g_\mathrm{NL}^{(\partial\sigma)^4}\rvert_\mathrm{eff}
	  = 0.0914 g_{\rm NL}^{\dot{\sigma}^2(\partial\sigma)^2}.
\end{equation}
These two relations determine how much the shape generated by $g_{\rm NL}^{\dot{\sigma}^2(\partial\sigma)^2}$ term overlaps with the shapes generated by $g_{\rm NL}^{\dot{\sigma}^4}$ and $g_{\rm NL}^{(\partial\sigma)^4}$.

Now, to impose the positivity bound in \cref{eq:LAM_bound}, we first rewrite it as
\be
	(8 - 8\gamma^2 + 3 \gamma^4) \frac{20736}{2575}g_{\rm NL}^{(\partial\sigma)^4} + (4\gamma^2 -3 \gamma^4) \frac{6912}{325}g_{\rm NL}^{\dot{\sigma}^2(\partial\sigma)^2} + \gamma^4 \frac{3\cdot768}{25}g_{\rm NL}^{\dot{\sigma}^4} > 0.
\ee
In particular, setting $\gamma=1$, i.e., working in the CM frame, we obtain
\be
\frac{3\cdot 20736}{2575} g_{\rm NL}^{(\partial\sigma)^4}  + \frac{6912}{325} g_{\rm NL}^{\dot{\sigma}^2(\partial\sigma)^2} + \frac{3\cdot 768}{25} g_{\rm NL}^{\dot{\sigma}^4} > 0.
\ee
Using the effective contributions in \cref{eq:eff_coeff}, we arrive at our final positivity bound, 
\bea
	\left[(8 - 8\gamma^2 + 3 \gamma^4) \frac{20736}{2575} +  0.0914(4\gamma^2 -3 \gamma^4) \frac{6912}{325} \right]g_{\rm NL}^{(\partial\sigma)^4} \\
	+ \left[\gamma^4 \frac{3\cdot768}{25} + 0.597(4\gamma^2 -3 \gamma^4) \frac{6912}{325} \right] g_{\rm NL}^{\dot{\sigma}^4} > 0.
\label{eq:positivityfinal}
\eea

\begin{figure}[h]
    \centering
    \includegraphics[width=\textwidth]{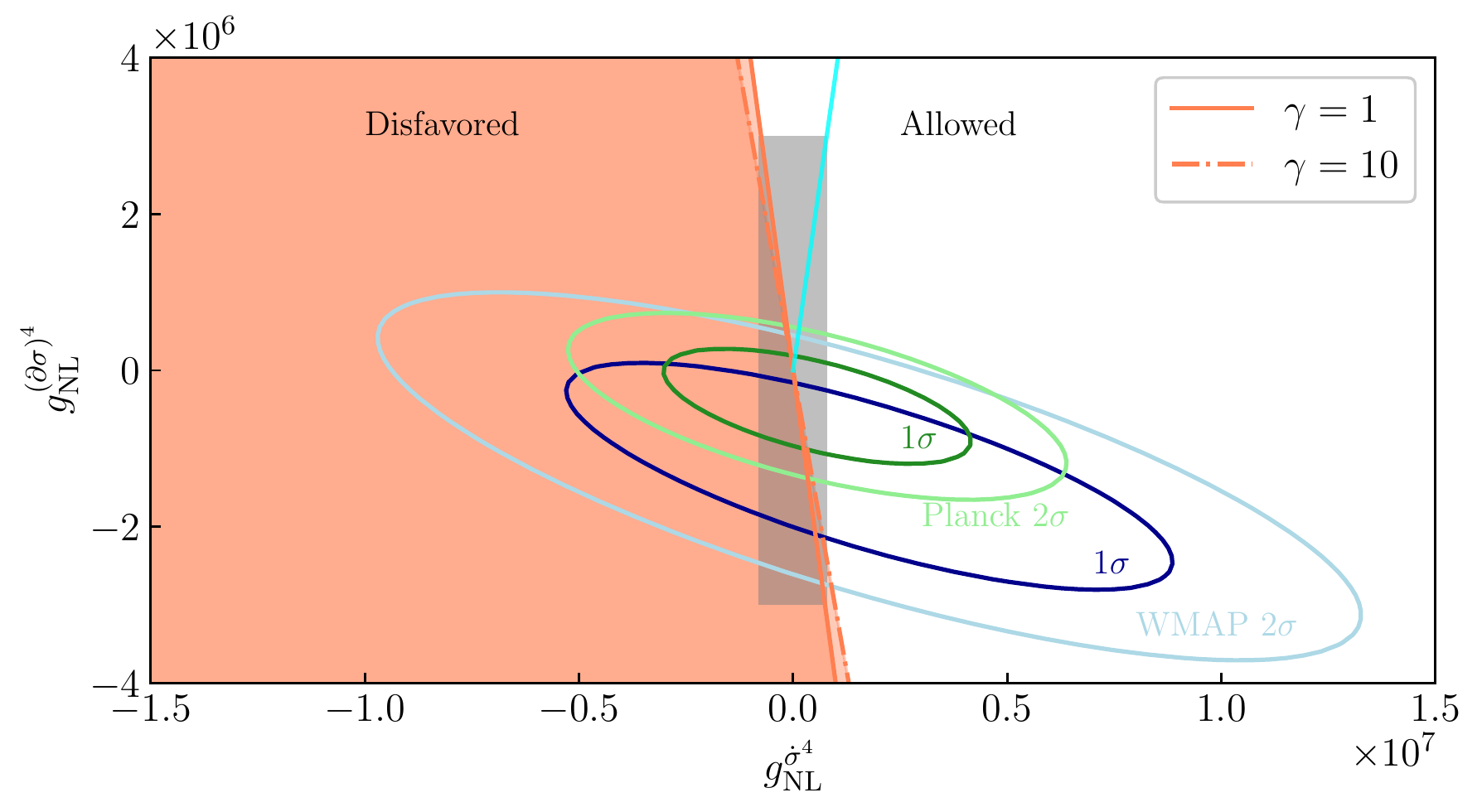}
    \vspace{-0.5cm}
    \caption{Positivity bounds on the trispectrum parameter space derived from unitarity, for representative values of the $\gamma$ parameter, compared with the region allowed by WMAP (blue) and Planck (green) constraints. The simple sigma-model UV completion discussed in text generates NG along the cyan line. 
    The gray rectangular region shows the parameter space for which $\Lambda_\chi > H$, as obtained from Eq.~\eqref{eq:gNLs} by setting various $e_i=1$.
    } 
    \label{fig:trispec_constraint}
\end{figure}

We illustrate the constraint of \cref{eq:positivityfinal} in \cref{fig:trispec_constraint}, where we take the WMAP $1\sigma$ and $2\sigma$ constraint from \Refc{Smith:2015uia} and use the Fisher matrices in \Refc{Planck:2019kim} to derive the corresponding bounds for the Planck data. We show the bounds for $\gamma = 1$ and $\gamma = 10$ in order to illustrate how this parameter space is affected by changing the Breit parameterization.
As mentioned above, we cannot raise $\gamma$ arbitrarily while retaining a valid EFT expansion. 
Even if one were to consider such a possibility, in the limit of $\gamma\gg 1$, the relation in \cref{eq:positivityfinal} becomes independent of $\gamma$ since all the coefficients scale as $\gamma^4$ and factor out of the bound.
Therefore, the result shown for $\gamma=10$ would continue to approximate the result for $\gamma\gg 1$.

Since applying positivity bounds require a hierarchy between $H$ and $\Lambda_\chi$, in Fig.~\ref{fig:trispec_constraint} we show the parameter space for which $\Lambda_\chi > H$.
To obtain this result we have used the expressions for $g_{\rm NL}^{\dot{\sigma}^4}$ and $g_{\rm NL}^{(\partial\sigma)^4}$ in Eq.~\eqref{eq:gNLs} and set $e_1=e_3=1$.
We see that while our positivity bound can meaningfully constrain parts of the parameter space where $\Lambda_\chi \gg H$, some parts of the Planck contours lie outside the gray region. 
With future improvements in the constraints on the trispectrum, these contours may eventually be completely contained inside the gray region and therefore be fully sensitive to scenarios with $\Lambda_\chi > H$.

We now discuss a simple UV completion where these dimension-eight operators arise with a Lorentz-invariant structure. Let us consider a complex scalar field $\Phi = (f_a + \rho/\sqrt{2})\exp[i a / (\sqrt{2}f_a)]$ whose radial mode is denoted by $\rho$ and the Goldstone mode by $a$.
Then the kinetic term for $\Phi$ gives rise to the following interactions:
\be
	|\partial\Phi|^2 = \frac{1}{2}(\partial \rho)^2 + \frac{1}{2}(\partial a)^2 + \frac{\rho}{\sqrt{2}f_a}(\partial a)^2 + \frac{\rho^2}{4 f_a^2}(\partial a)^2.
\ee
Adding a potential $V(\Phi)=m_\rho^2 (|\Phi|^2 - f_a^2)^2/4f_a^2$ generates a mass $m_\rho$ for $\rho$ with no tadpole.
Integrating out the radial mode $\rho$ at tree level generates a dimension-eight operator~\cite{Adams:2006sv},
\be
	\frac{1}{4f_a^2 m_\rho^2}(\partial a)^4.
\ee
Therefore, for this Lorentz-invariant example,  $e_1 = e_3 = -e_2/2 > 0$ in the notation of \cref{eq:LAM} with $a$ replaced by $\chi$. In terms of the $g_{\rm NL}$ coefficients, this implies $g_{\rm NL}^{(\dot{\sigma})^4} =(27/103) g_{\rm NL}^{(\partial\sigma)^4} > 0$.
This is shown via the cyan line in \cref{fig:trispec_constraint}.

\subsection{Constraints on $\mathcal{L}_\mathrm{UV}$}

We now focus on a two-field scenario where both adiabatic and isocurvature perturbations arise.
We will connect our positivity bounds to observational constraints involving both adiabatic and isocurvature perturbations, which we denote by $\zeta$ and $S$, respectively.
While this scenario can be realized using $\mathcal{L}_{\rm IM}$ in \cref{eq:L_IM}, we choose to focus on $\mathcal{L}_{\rm UV}$ in \cref{eq:brokenUVEFT} for simplicity.

We first use the results of \cref{app:bispec_comp} to compute the various three-point functions involving $\zeta$ and $S$. 
For this purpose, we need to relate the field fluctuations of the inflaton $\xi$ and the spectator field $\delta\sigma$ to $\zeta$ and $S$, respectively.
We note that here we refer to the spectator field as $\delta\sigma$ rather than $\sigma$ as earlier, as now we expand the spectator about a misaligned value $\sigma_0 \gg \delta\sigma$, which it obtains at the time of horizon exit of $\delta\sigma$.

We assume that the inflaton dominates the energy density during inflation and sources a superhorizon curvature perturbation $\zeta \simeq - H\xi/\dot{\phi}_0$, where as before $H$ is the Hubble scale during inflation and $|\dot{\phi}_0|^{1/2}$ is the slow-roll velocity of the inflaton. 
Here have used the spatially flat gauge~\cite{Maldacena:2002vr} to relate the inflaton perturbation to the curvature perturbation.
The adiabatic power spectrum is then given by
\be
  \ev{\zeta(\vk)\zeta(-\vk)}' =  P_\zeta(k) = \frac{H^4}{\dot{\phi}_0^2}\frac{1}{2k^3}.
\ee
The measured CMB anisotropies fix $|\dot{\phi}_0|^{1/2} \simeq 60 H$~\cite{Planck:2018jri}.

While the energy density in the spectator field $\sigma$ is subdominant during inflation, it acquires (almost) scale-invariant isocurvature fluctuations.
For concreteness, we assume $\sigma$ sources dark matter perturbations at later times, and consequently such fluctuations manifest as uncorrelated dark matter isocurvature fluctuations.
In terms of $S$ and at the linear order, these can be parameterized as $S = 2 \delta\sigma/\sigma_0$.
The isocurvature power spectrum is then given by
\begin{equation}
  \ev{S(\vk)S(-\vk)}' = P_S(k) = \frac{4 H^2}{\sigma_0^2} \frac{1}{2k^3}.
\end{equation}
For simplicity, in the following we will assume that the spectral tilt of $P_S(k)$ is the same as that of $P_\zeta(k)$.
With these relations at hand, we can compute various three-point functions, which are explicitly described in \cref{app:bispec_comp}.
In particular, we can use \cref{eq:xi3,eq:xi2sig,eq:xisig2,eq:sig3} and define  $k_{ij \dotsm m} = k_i + k_j + \dotsb + k_m$ to write   
\begin{equation}
\label{eq:zeta3}
  \begin{split}
	&\ev{\zeta(\vk_1) \zeta(\vk_2) \zeta(\vk_3)}' \\
	&\qquad = -\frac{2\kappa_1 H^{8}}{\Lambda^4\dot{\phi}_0^2 k_1^3 k_2^3 k_3^3}
	          \bqty{\frac{2k_1^2 k_2^2 k_3^2}{k_{123}^3}
	                - \frac{(\vk_2\cdot\vk_3) k_1^2}{k_{123}}
	       	          \pqty{1 + \frac{k_{23}}{k_{123}} + \frac{2k_2 k_3}{k_{123}^2}}}
           	   + \text{2 perms.,}
           	   \end{split}
           	   \end{equation}
\begin{equation}
\label{eq:zeta2S}
\begin{split}
	&\ev{S(\vk_1) \zeta(\vk_2) \zeta(\vk_3)}' \\
	&\qquad = \frac{\kappa_2 H^7}{2\Lambda^4\sigma_0 \dot{\phi}_0 k_1^3 k_2^3 k_3^3}
	          \bqty{\frac{2k_1^2 k_2^2 k_3^2}{k_{123}^3}
	          	    - \frac{(\vk_2\cdot\vk_3) k_1^2}{k_{123}}
	          	      \pqty{1 + \frac{k_{23}}{k_{123}} + \frac{2k_2 k_3}{k_{123}^2}} }\\
	&\qquad\qquad + \frac{\kappa_2 H^7}{2\Lambda^4\sigma_0 \dot{\phi}_0 k_1^3 k_2^3 k_3^3}
	                \Bigg\{\bqty{\frac{2k_1^2 k_2^2 k_3^2}{k_{123}^3}
	                		    - \frac{(\vk_2\cdot\vk_1) k_3^2}{k_{123}}
	                		      \pqty{1 + \frac{k_{12}}{k_{123}} + \frac{2k_1 k_2}{k_{123}^2}}} \\& \hspace{112mm}
                          + (\vk_2\leftrightarrow \vk_3)\Bigg\},
\end{split}
\end{equation}
\begin{equation}
\label{eq:zetaS2}
\begin{split}
	&\ev{S(\vk_1) S(\vk_2) \zeta(\vk_3)}' \\
	&\qquad = -\frac{\kappa_3 H^6}{\Lambda^4\sigma_0^2 k_1^3 k_2^3 k_3^3}
	          \bqty{\frac{2k_1^2 k_2^2 k_3^2}{k_{123}^3}
	          	    - \frac{(\vk_1\cdot \vk_2) k_3^2}{k_{123}}
	          	      \pqty{1 + \frac{k_{12}}{k_{123}} + \frac{2k_1 k_2}{k_{123}^2}} }\\
	&\qquad\qquad - \frac{\kappa_4 H^6}{2\Lambda^4\sigma_0^2 k_1^3 k_2^3 k_3^3}
	                \Bigg\{\bqty{\frac{2k_1^2 k_2^2 k_3^2}{k_{123}^3}
	                		    - \frac{(\vk_1\cdot \vk_3) k_2^2}{k_{123}}
	                		      \pqty{1 + \frac{k_{13}}{k_{123}} + \frac{2k_1 k_3}{k_{123}^2}}} \\& \hspace{112mm}
                          + (\vk_1\leftrightarrow \vk_2)\Bigg\},
                          \end{split}
                          \end{equation}
and
\begin{equation}
\label{eq:S3}
\begin{split}
	&\ev{S(\vk_1) S(\vk_2) S(\vk_3)}' \\
	&\qquad =  \frac{\kappa_5\phid H^5}{2\Lambda^4\sigma_0^3 k_1^3 k_2^3 k_3^3}
	           \bqty{\frac{2k_1^2 k_2^2 k_3^2}{k_{123}^3}
	           	     - \frac{(\vk_2\cdot \vk_3) k_1^2}{k_{123}}
	           	       \pqty{1 + \frac{k_{23}}{k_{123}} + \frac{2k_2 k_3}{k_{123}^2}}} 
           	   + \text{2 perms.}
  \end{split}
\end{equation}
Since these three-point functions originate from derivative interactions in \cref{eq:brokenUVEFT}, the resulting NG is generated during the horizon-crossing times of the modes.
Consequently for comparable $\kappa_i$, the above three-point functions peak at the ``equilateral'' momentum configuration $k_1 \simeq k_2 \simeq k_3$.

To compare, we consider the Planck search for isocurvature NG, which is based on the templates~\cite{Langlois:2012tm},
\begin{equation}
\label{eq:planck_3pt}
  \begin{split} 
   &\ev*{X^I(\vk_1) X^J(\vk_2) X^K(\vk_3)}'\\
   &\qquad = \tilde{f}_\mathrm{NL}^{I,JK} P_\zeta(k_2) P_\zeta(k_3)
      + \tilde{f}_\mathrm{NL}^{J,KI} P_\zeta(k_1) P_\zeta(k_3)
      + \tilde{f}_\mathrm{NL}^{K,IJ} P_\zeta(k_1) P_\zeta(k_2).
  \end{split}
\end{equation}
The indices $I,J,K$ each can denote an adiabatic ($a$) or an isocurvature fluctuation ($i$), \ie, $I, J, K = a, i$ with $X^a = \zeta$ and $X^i = S$.
While the Planck collaboration has constrained the parameters $\tilde{f}_\mathrm{NL}^{\zeta,\zeta\zeta}, \tilde{f}_\mathrm{NL}^{\zeta, \zeta S}, \tilde{f}_\mathrm{NL}^{S, \zeta\zeta}, \tilde{f}_\mathrm{NL}^{\zeta, SS}, \tilde{f}_\mathrm{NL}^{S, \zeta S}, \tilde{f}_\mathrm{NL}^{S ,SS}$, we cannot directly use those bounds in the present context.
This is because the associated three-point functions in \cref{eq:planck_3pt} peak at the ``squeezed'' momentum configuration, where one momentum is much smaller than the other two, $k_3 \ll k_1, k_2$.
Therefore, the template in \cref{eq:planck_3pt} is not well suited to constrain the shapes given by \cref{eq:zeta3,eq:zeta2S,eq:zetaS2,eq:S3}, and a dedicated analysis would be more appropriate.

Given this obstacle, we here take a simple approach to illustrate the positivity bounds.
For the adiabatic-only scenario, we can evaluate $\langle \zeta(\vec{k}_1) \zeta(\vec{k}_2) \zeta(\vec{k}_3) \rangle'$ in the equilateral limit and demand that it obeys the bounds of equilateral NG parameter $f_{\rm NL}^{\rm equi}$ constrained by Planck~\cite{Planck:2019kim}.
Extending that condition to include other perturbations, we define the following quantities, keeping  \cref{eq:zeta3,eq:zeta2S,eq:zetaS2,eq:S3} in mind:
\bea
	f_{\rm NL,equi}^{aaa} & = \frac{\langle \zeta(\vec{k}) \zeta(\vec{k}) \zeta(\vec{k}) \rangle'}{P_\zeta(k)^2}\\
	f_{\rm NL,equi}^{aai} & = \frac{\langle \zeta(\vec{k}) \zeta(\vec{k}) S(\vec{k}) \rangle'}{P_\zeta(k)^2}\\
	f_{\rm NL,equi}^{aii} & = \frac{\langle \zeta(\vec{k}) S(\vec{k}) S(\vec{k}) \rangle'}{P_\zeta(k)^2}\\
	f_{\rm NL,equi}^{iii} & = \frac{\langle S(\vec{k}) S(\vec{k}) S(\vec{k}) \rangle'}{P_\zeta(k)^2}.
\eea
We consider a benchmark of $P_S(k) = 10^{-2} P_\zeta(k)$, consistent with current searches for uncorrelated dark matter isocurvature~\cite{Planck:2018jri}.
This corresponds to $H/\sigma_0 \simeq 1.5\times 10^{-5}$.
Then the above set of $f_{\rm NL,equi}$ can be computed as
\bea
		f_{\rm NL, equi}^{aaa} & = -\frac{28}{3}\frac{\kappa_1 H^4}{\Lambda^4} \frac{\dot{\phi}_0^2}{H^4} \simeq -1.2 \times 10^8 \cdot \frac{\kappa_1 H^4}{\Lambda^4}\\
		f_{\rm NL, equi}^{aai} & = \frac{7}{3} \frac{\kappa_2 H^4}{\Lambda^4} \frac{H}{\sigma_0} \frac{\dot{\phi}_0^3}{H^6} \simeq  1.6 \times 10^6 \cdot \frac{\kappa_2 H^4}{\Lambda^4}\\
		f_{\rm NL, equi}^{aii} & = -\frac{14}{9}\frac{(\kappa_3+\kappa_4)H^4}{\Lambda^4}\frac{\dot{\phi}_0^4}{H^8}\frac{H^2}{\sigma_0^2} \simeq -5.8\times 10^4 \cdot \frac{(\kappa_3+\kappa_4)H^4}{\Lambda^4}\\
		f_{\rm NL, equi}^{iii} & = \frac{7}{3}\frac{\kappa_5 H^4}{\Lambda^4}\frac{\dot{\phi}_0^5}{H^{10}}\frac{H^3}{\sigma_0^3} \simeq 4.7\times 10^3\frac{\kappa_5 H^4}{\Lambda^4}.
\eea
With a dedicated search for the NG shapes described in \cref{eq:zeta3,eq:zeta2S,eq:zetaS2,eq:S3}, one can bound the set of $f_\mathrm{NL, equi}$ parameters and consequently, the various coefficients $\kappa_i$.
The bound from the generalized optical theorem given in \cref{eq:UVbound_gen_opt} can then be used to further constrain the same parameter space.

\section{Conclusions}\label{sec:concl}

Positivity bounds represent nontrivial restrictions on EFTs and can dictate which effective actions can be UV-completed into a Lorentz-invariant, causal, unitary theory.
An important ingredient in the derivation of such positivity bounds is analyticity of the S-matrix, which follows from causality in Lorentz-invariant EFTs.
This is precisely one reason why the application of positivity bounds in theories where Lorentz invariance is spontaneously broken is subtle.

In this work, we have focused on one such class of EFTs, namely, those that can describe the inflationary Universe.
During inflation, time translation symmetry is spontaneously broken, and at energies below the symmetry-breaking scale, the EFT typically would not have a manifestly Lorentz-invariant description.
Consequently, fluctuations in the EFT can propagate at speeds different from the speed of light.
More importantly, if there are multiple propagating fluctuating modes, some will generically propagate faster than the rest.
This case can then give rise to scenarios where the faster-moving fluctuations mediate interactions between the slower-moving modes that are apparently ``acausal'' from the perspective of the slow degrees of freedom.
We have seen this phenomenon happen explicitly when working in the Breit parameterization in a two-field scenario where such interactions can give rise to poles of the S-matrix away from the real $s$ axis, invalidating the standard prescription for deriving positivity bounds using analyticity.

Given this pathology, we have focused on multifield scenarios where all the fluctuations propagate at speeds parametrically close to the speed of light.
While in such scenarios the S-matrix is analytic, one can still have nontrivial restrictions on theories originating from Lorentz-breaking effects.
To illustrate these bounds, we have focused on four types of inflationary EFTs, including both adiabatic and isocurvature perturbations.
By applying conventional elastic bounds as well as more stringent constraints from the generalized optical theorem, all the while employing the Breit parameterization, we have obtained new positivity bounds on various Wilson coefficients of the inflationary EFTs.

Interestingly, these coefficients also give rise to NG of primordial perturbations that can be searched for using CMB and large-scale structure observations.
Therefore, our positivity bounds can nontrivially constrain the parameter space relevant for such searches.
As an illustration, we have focused on a Planck trispectrum search with adiabatic fluctuations and demonstrated how positivity bounds can disfavor large parts of the parameter space that is otherwise viable.
Similar conclusions can be obtained in the context of NG searches when both adiabatic and isocurvature perturbations are present.
While the current Planck constraints from isocurvature NG are not directly applicable to our predicted shape of the bispectrum, we have computed such shapes and derived the associated positivity bounds.
Along the way, we have also shown how, in a multifield scenario, bounds from the generalized optical theorem can give a stronger constraints over standard elastic positivity bounds alone.

Looking forward, arguably the most pressing open question is to determine whether positivity arguments can be constructed in theories with multiple distinct speeds of sound.
Resolving this issue is not merely a technical requirement for multifield EFTs, but rather a challenge for studying nonrelativistic EFTs more generally.
Even for single-field theories, if $c_s < 1$, the UV can introduce additional states propagating outside this cone, and the apparent acausality returns, modifying the analytic properties of UV amplitudes and thereby undermining the dispersive arguments that lead to positivity when working in the Breit parameterization.
Whatever form of modification to the conventional approach is required to achieve this generalization, the result would open the power of positivity to the enormous range of EFTs that do not exhibit Lorentz invariance.

\vspace{5mm}
 
\begin{center} 
{\bf Acknowledgments}
\end{center}
\noindent 
We thank Simone Ferraro, Daniel Green, and Scott Melville for useful discussions and comments.
M.F.~is supported by the U.S. Department of Energy (DOE) under grant DE-SC0010008.
S.K.~is supported in part by the U.S. National Science Foundation (NSF) grant PHY-1915314 and the DOE contract DE-AC02-05CH11231. 
M.F. \& S.K. thank the Mainz Institute of Theoretical Physics of the Cluster of Excellence PRISMA+ (Project ID 39083149) for its hospitality while this work was in progress.
M.F. also thanks the Aspen Center for Physics, supported by NSF grant PHY-1607611, for its hospitality while this work was in progress.
G.N.R. is supported at the Kavli Institute for Theoretical Physics by the Simons Foundation (Grant No. 216179) and the NSF (Grant No. PHY-1748958) and at the University of California, Santa Barbara by the Fundamental Physics Fellowship.

\begin{appendix}
\section{Computation of Bispectrum Shapes}
\label{app:bispec_comp}

In this appendix, we present various details associated with the computation of various three-point functions in the context of the Lagrangian in \cref{eq:brokenUVEFT}. 
We follow the notation in Ref.~\cite{Kumar:2017ecc} and refer the reader to Ref.~\cite{Chen:2010xka} for a review of primordial non-Gaussianity.

We treat both $\xi$ and $\sigma$ as massless fields during inflation.
To write down their inflationary mode functions, we first parameterize the inflationary spacetime metric in conformal coordinates as
\be
  \dd{s^2} = \frac{1}{\eta^2 H^2} (-\dd{\eta^2} + \dd{\vec{x}^2}).
\ee
Here $H$ is the Hubble scale during inflation, which we take to be approximately constant.
For simplicity, we will now work in units where $H=1$ and later reintroduce $H$ by dimensional analysis.
The mode functions for massless scalar fields follow after canonical quantization, and we can write them as
\be
  \Phi(\eta, \vk) = f_k(\eta) a_{\vk}^\dagger + \bar{f}_k(\eta) a_{-\vk},
\ee
where $\Phi$ can be either $\xi$ or $\sigma$, and $k$ is the comoving momentum of the fluctuating mode.
The mode functions $f_k(\eta)$ are given by
\be
  \begin{split}
    f_k(\eta) &= \frac{1}{\sqrt{2k^3}} (1-ik\eta) e^{ik\eta},\\
    \bar{f}_k(\eta) &= \frac{1}{\sqrt{2k^3}} (1+ik\eta) e^{-ik\eta}.
  \end{split}
\ee

Using these mode functions, we can compute the relevant three-point functions using the ``in-in'' formalism (see, \eg, \Refc{Weinberg:2005vy}).
This computation determines the expectation value of an observable $\oper(t)$ at time $t_f$ as
\be
\label{eq:inin}
	\langle \oper(t_f) \rangle \left.\equiv
	\langle 0| U(t_f, t_i)^\dagger \oper_I(t_f) U(t_f, t_i)
	|0\rangle\right|_{t_i \to -\infty(1-i\epsilon)}
\ee
where
\be
	U(t_f, t_i) = \mathrm{T} \exp(-i\int_{t_i}^{t_f} \dd{t} H_I^\mathrm{int}(t)),
\ee
is the time evolution operator with $H_I^\mathrm{int}(t)$ the interacting part of the Hamiltonian evaluated in the interaction picture.
Thus, the cosmological correlation functions involve both the time-ordering operator $\mathrm{T}$ and the anti-time-ordering operator $\bar{\mathrm{T}}$ (via $U^\dagger$), contrary to a computation of the S-matrix.
Other than this difference, the computation of expectation values can be carried out by expanding the exponentials in \cref{eq:inin} along with Wick contractions, as in a standard flat-space computation.
In particular, we take the operator ${\cal O}_I(t_f)$ in Eq.~\eqref{eq:inin} to be ${\cal O}_I(t_f) \equiv \Phi(\eta_0\rightarrow 0, \vec{k}_1) \Phi(\eta_0\rightarrow 0, \vec{k}_2) \Phi(\eta_0\rightarrow 0, \vec{k}_3)$ since we are interested in a computation of the three point function.
Here $\eta_0$ denotes a conformal time towards the end of inflation, corresponding to $t_f$ in~\cref{eq:inin}, where the expectation value is evaluated.

Looking at \cref{eq:inin,eq:brokenUVEFT}, we observe that to compute the three-point functions at leading order in couplings and at tree level, we need to calculate only the time-ordered contribution.
We can then obtain the full answer by adding its complex conjugate, corresponding to the anti-time-ordered piece.
Furthermore, \cref{eq:brokenUVEFT} has both temporal and spatial derivatives acting on the fluctuations.
Therefore, we need only two types of Wick contractions (at the leading order in perturbation theory), 
\bea
\ev*{\Phi(\eta_0 \to 0, \vk) \dot{\Phi}(\eta, -\vk)}
                      &= -\frac{\eta^2 k^2}{2k^3} e^{ik\eta}\\
\ev{\Phi(\eta_0 \to 0, \vec{k}) \partial_i{\Phi}(\eta, -\vk)}
                      &= \frac{1}{2k^3}(-ik_i)(1-ik\eta) e^{i k\eta},
\eea
from which we can build the rest of the correlator.
We also rewrite here the first line of \cref{eq:brokenUVEFT} for convenience,
\begin{equation}
\label{eq:brokenUVEFT_3pt}
  \mathcal{L}_\mathrm{UV} \supset
    \frac{\phid}{\Lambda^4}
      \bqty{-(4\kappa_1\dot{\xi} + \kappa_2 \sigd)(\partial\xi)^2
	        - 2(\kappa_2\dot{\xi} + \kappa_4 \sigd)(\partial^\mu\xi\partial_\mu\sigma)
	        - (2\kappa_3\dot{\xi} + \kappa_5\sigd)(\partial\sigma)^2}.
\end{equation}

\paragraph{\boldmath $\ev{\xi\xi\xi}$ correlator.}
The inflaton three-point function receives a contribution from the $\kappa_1$ term in \cref{eq:brokenUVEFT_3pt} and is given by
\begin{equation}
  \begin{split}
	& \ev{\xi(\vk_1) \xi(\vk_2) \xi(\vk_3)} \\
	&\qquad =  -\frac{2\cdot 4 i \kappa_1}{8k_1^3 k_2^3 k_3^3}\frac{\phid}{\Lambda^4}
	              \int_{-\infty}^0 \frac{\dd{\eta}}{\eta^4} (\eta^2 k_1^2)
	              \bqty{k_2^2 k_3^2 \eta^4 +  \eta^2 (\vk_2\cdot \vk_3)(1- i k_{23}\eta - k_2 k_3 \eta^2) } e^{ik_{123}\eta}  \\
	&\qquad\qquad  + \text{c.c.} + \text{perms.}
  \end{split}
\end{equation}
As in the main text, we have denoted $k_{ij \dotsm m} = k_i + k_j + \dotsb + k_m$.
To compute the above integral, we can use the relation,
\be
  \int_{-\infty}^0 \dd{\eta} \eta^\alpha e^{ik\eta}
    = -\frac{i^{1 + \alpha}}{k^{1 + \alpha}}\, \Gamma(1 + \alpha).
\ee
After further simplification, this leads to
\begin{equation}
\label{eq:xi3}
  \begin{split}
	&\ev{\xi(\vk_1) \xi(\vk_2) \xi(\vk_3)} \\
	&\qquad = \frac{2\kappa_1}{k_1^3 k_2^3 k_3^3}\frac{\phid}{\Lambda^4}
	          \bqty{\frac{2k_1^2 k_2^2 k_3^2}{k_{123}^3}
	          	    - \frac{(\vk_2\cdot \vk_3) k_1^2}{k_{123}}
	          	      \pqty{1 + \frac{k_{23}}{k_{123}} + \frac{2k_2 k_3}{k_{123}^2} }}
	          + \text{2 perms.}
  \end{split}
\end{equation}

\paragraph{\boldmath $\langle \sigma \xi \xi \rangle$ correlator.} 
This correlator has the same structure as above since both $\sigma$ and $\xi$ are massless fields, except we need to account for the two terms proportional to $\kappa_2$ in \cref{eq:brokenUVEFT_3pt}. 
The result is still given by
\begin{equation}
\label{eq:xi2sig}
  \begin{split}
	&\ev{\sigma(\vk_1) \xi(\vk_2) \xi(\vk_3)} \\
	&\qquad = \frac{\kappa_2}{2k_1^3 k_2^3 k_3^3}\frac{\phid}{\Lambda^4}
	          \bqty{\frac{2k_1^2 k_2^2 k_3^2}{k_{123}^3}
	          	    - \frac{(\vk_2\cdot \vk_3) k_1^2}{k_{123}}
	          	      \pqty{1 + \frac{k_{23}}{k_{123}} + \frac{2k_2 k_3}{k_{123}^2}} } \\
	&\qquad\qquad + \Bqty{\frac{\kappa_2}{2k_1^3 k_2^3 k_3^3}\frac{\phid}{\Lambda^4}
		                  \bqty{\frac{2k_1^2 k_2^2 k_3^2}{k_{123}^3}
		                  	    - \frac{(\vk_2 \cdot \vk_1) k_3^2}{k_{123}}
		                  	            \pqty{1 + \frac{k_{12}}{k_{123}}
		                  	            	  + \frac{2k_1 k_2}{k_{123}^2}} }
	                  	  + (\vk_2\leftrightarrow \vk_3)}.
  \end{split}
\end{equation}

\paragraph{\boldmath $\langle \sigma \sigma \xi \rangle$ correlator.} 
The structure remains same as above, except the individual contributions are controlled by $\kappa_3$ and $\kappa_4$:
\begin{equation}
\label{eq:xisig2}
  \begin{split}
	&\ev{\sigma(\vk_1) \sigma(\vk_2) \xi(\vk_3)} \\
	&\qquad = \frac{\kappa_3}{k_1^3 k_2^3 k_3^3}\frac{\phid}{\Lambda^4}
	          \bqty{\frac{2k_1^2 k_2^2 k_3^2}{k_{123}^3}
	          	    - \frac{(\vk_1 \cdot \vk_2) k_3^2}{k_{123}}
	          	        \pqty{ 1 + \frac{k_{12}}{k_{123}} + \frac{2k_1 k_2}{k_{123}^2}} } \\
    &\qquad\qquad + \Bqty{\frac{\kappa_4}{2k_1^3 k_2^3 k_3^3}\frac{\phid}{\Lambda^4}
    	                    \bqty{\frac{2k_1^2 k_2^2 k_3^2}{k_{123}^3}
    	                    	  - \frac{(\vk_1\cdot \vk_3) k_2^2}{k_{123}}
    	                    	      \pqty{1 + \frac{k_{13}}{k_{123}} + \frac{2k_1 k_3}{k_{123}^2}} }
                        	      + (\vk_1 \leftrightarrow \vk_2) }.
  \end{split}
\end{equation}

\paragraph{\boldmath $\langle \sigma \sigma \sigma \rangle$ correlator.} 
This case is identical to the $\langle\xi\xi\xi\rangle$ correlator, except that $4\kappa_1$ is replaced by $\kappa_5$:
\begin{equation}
\label{eq:sig3}
  \begin{split}
	&\ev{\sigma(\vk_1) \sigma(\vk_2) \sigma(\vk_3)} \\
	&\qquad = \frac{\kappa_5}{2k_1^3 k_2^3 k_3^3}\frac{\phid}{\Lambda^4}
	          \bqty{\frac{2k_1^2 k_2^2 k_3^2}{k_{123}^3}
	          	    - \frac{(\vk_2\cdot \vk_3) k_1^2}{k_{123}}
	          	      \pqty{1 + \frac{k_{23}}{k_{123}} + \frac{2k_2 k_3}{k_{123}^2}} }
          	  + \text{2 perms.}
  \end{split}
\end{equation}

\end{appendix}
\bibliographystyle{JHEP}
\bibliography{references}
\end{document}